\documentclass[acmsmall]{acmart}

\def\BibTeX{{\rm B\kern-.05em{\sc i\kern-.025em b}\kern-.08emT\kern-.1667em\lower.7ex\hbox{E}\kern-.125emX}}
    
\setcopyright{acmlicensed}
\acmJournal{POMACS}
\acmYear{2019} \acmVolume{3} \acmNumber{1} \acmArticle{6} \acmMonth{1} \acmPrice{15.00}\acmDOI{10.1145/3311077}

\usepackage{booktabs}
\usepackage{amsmath,amssymb}
\usepackage{bbm}
\usepackage{bm}
\usepackage{subfigure}

\newtheorem{theorem}{Theorem}[section]

\DeclareMathOperator*{\argmax}{\arg\max}
\DeclareMathOperator*{\argmin}{\arg\min}

\begin{document}

%
\title[Location-Based Advertising]{Analyzing Location-Based Advertising for Vehicle\\Service Providers Using Effective Resistances}

%
\author{Haoran Yu}
\email{haoran.yu@northwestern.edu}
\author{Ermin Wei}
\email{ermin.wei@northwestern.edu}
\author{Randall A. Berry}
\email{rberry@eecs.northwestern.edu}
\affiliation{%
  \institution{Department of Electrical and Computer Engineering, Northwestern University}
  \streetaddress{2145 Sheridan Road}
  \city{Evanston}
  \state{Illinois}
  \postcode{60208}
}

%
\renewcommand{\shortauthors}{H. Yu, E. Wei, and R. A. Berry}

%
\begin{abstract}
Vehicle service providers can display commercial ads in their vehicles based on passengers' origins and destinations to create a new revenue stream. In this work, we study a vehicle service provider who can generate different ad revenues when displaying ads on different arcs (i.e., origin-destination pairs). The provider needs to ensure the vehicle flow balance at each location, which makes it challenging to analyze the provider's vehicle assignment and pricing decisions for different arcs. For example, the provider's price for its service on an arc depends on the ad revenues on other arcs as well as on the arc in question. To tackle the problem, we show that the traffic network corresponds to an electrical network. When the \emph{effective resistance} between two locations is small, there are many paths between the two locations and the provider can easily route vehicles between them. We characterize the dependence of an arc's optimal price on any other arc's ad revenue using the effective resistances between these two arcs' origins and destinations. Furthermore, we study the provider's optimal selection of advertisers when it can only display ads for a limited number of advertisers. If each advertiser has one target arc for advertising, the provider should display ads for the advertiser whose target arc has a small effective resistance. {{We investigate the performance of our advertiser selection strategy based on a real-world dataset.}}
\end{abstract}

%
%


%

%
\maketitle

\section{Introduction}
In-vehicle advertising has been emerging as a promising approach for vehicle service providers (e.g., taxi companies and ride-sharing platforms) to monetize their service. Vehicle service providers can display ads to users of their service, and receive payment from the corresponding advertisers. Recently, there have been many companies (e.g., VUGO \cite{Vugo}, VIEWDIFY \cite{viewdify}, and SURF \cite{SURF}) that offer technical support for the in-vehicle advertising. By installing these companies' tablets in the vehicles, the vehicle service providers can provide various entertainment options (e.g., watching TV shows and playing games) to users and improve their in-vehicle experience. Meanwhile, the providers can deliver ads (including banner ads and in-stream ads) via the tablets.

The in-vehicle advertising enables advertisers to target users based on their travels' origins and destinations. For example, many destinations of vehicle service users are in shopping areas \cite{Vugo}, and the advertisers can advertise relevant promotional activities to these users. Likewise, the advertisers can direct the users whose destinations are in residential areas to online orders for food delivery \cite{fooddelivery}. {{VUGO analyzed around 30,000 rides and showed the diversity of users' destinations \cite{VugoData}. The company designs algorithms for vehicle service providers to achieve the targeted advertising.}}

With the advent of autonomous vehicles, in-vehicle advertising is likely to become increasingly common and powerful \cite{livingroom,fooddelivery}. The autonomous-driving technology will turn drivers into passengers, and vehicles will become ``moving living rooms'' \cite{livingroom}. Passengers can interact with various in-vehicle devices (e.g., touchscreens and holographic projectors), which are effective channels for advertising.

\subsection{Our Work}
In this work, we study the economic impact of in-vehicle advertising on a vehicle service provider who displays banner ads or in-stream ads to its users regularly (e.g., every 10 minutes). We use ``an arc'' to represent an origin-destination pair, and a ``unit ad revenue'' to represent the revenue that the provider can gain by displaying ads to a user for a time slot (e.g., the length of a time slot can be 10 minutes).{\footnote{We use an example to estimate the value of the unit ad revenue. According to \cite{YoutubeReport}, the average cost per click (CPC) of Youtube ads was around \$$3.58$ in the second quarter of 2018. {{Based on \cite{OKsource} and \cite{platformsource2}, the click-through rate (CTR) of an in-stream video ad was around $1.84$\% (note that this number was achieved without location-based targeting and the actual CTR of an in-vehicle video ad may be higher). When a provider displays $3$ in-stream video ads to a user during each time slot (10 minutes), the unit ad revenue is \$$0.20$ ($=3\times 1.84\%  \times \$3.58$).}} This value is comparable to a vehicle service provider's net profit when it does not display ads. According to \cite{Uberprofit2}, Uber's net profit is around \$8.23 per hour per vehicle, which corresponds to \$1.37 per time slot per vehicle.}} The advertisers have different target arcs to display their ads, and also have different willingnesses to pay. Hence, the provider's unit ad revenues on different arcs are different. 

The provider can set different prices for the vehicle service on different arcs \cite{bimpikis2016spatial,waserhole2012vehicle,ma2018spatio}. The key questions that we aim to answer are as follows:

\emph{Q1: Given the unit ad revenues, what are the provider's optimal service prices for different arcs}?

\emph{Q2: How do the provider's payoff and consumer surplus change with the unit ad revenues}?{\footnote{{{When a user takes the vehicle service, the corresponding \emph{consumer surplus} is the difference between the highest price that the user is willing to pay for the service and the price that it actually pays.}}}}

\emph{Q3: If the provider can only collaborate with a limited number of advertisers (who have different target arcs to display ads), which advertisers should the provider collaborate with}?

Answering these questions can help us understand the economic impact of in-vehicle advertising and provide insights for optimizing the provider's strategies in the presence of in-vehicle advertising. However, it is challenging to answer these questions, since the provider's vehicle assignment and pricing decisions for different arcs (i.e., origin-destination pairs) are tightly coupled by the \emph{vehicle flow balance constraints}. In the system's steady state, the mass of vehicles that depart from a location should equal the mass of vehicles that arrive at the location. Because of the flow balance constraints, the provider's vehicle assignment and pricing for one arc are affected by the network topology and the other arcs' parameters (e.g., traffic demand, travel time, and unit ad revenues). The flow balance constraints do not exist in other advertising models (e.g., web advertising \cite{ghosh2015match} and mobile advertising \cite{yu2019rewards}), and constitute a unique challenge to the study of in-vehicle advertising.

It is difficult to directly derive the provider's optimal decisions and payoff as the closed-form expressions of system parameters. Our key idea is to show that the traffic network, which is determined by the network topology and the traffic demand and travel time on different arcs, has an associated electrical network.{\footnote{The idea is inspired by our observation that the optimal dual variables associated with the flow balance constraints are the solutions to a system of linear equations, whose coefficient matrix is a Laplacian matrix and has a strong connection with an electrical network.}} Each location corresponds to a node in the electrical network. If there is a positive traffic demand between two locations, there is a resistor between the two corresponding nodes, and its resistance is determined by the traffic demand and travel time between the two locations. Given the associated electrical network, we can compute the \emph{effective resistance} between any two nodes. Intuitively, a small effective resistance implies that there are many paths between the two locations and the provider can easily route vehicles between them. Based on the effective resistances, we can derive closed-form expressions for the provider's optimal decisions and payoff, and further answer the three key questions. 


We summarize our answers to the three questions as follows.

{\bf{1. Provider's Optimal Spatial Pricing (Theorems \ref{theorem:priceresistance} and \ref{theorem:adarc}).}} Given the unit ad revenues, the provider's optimal price for an arc $\left(i,j\right)$ consists of two components. The first component is only related to the arc's unit ad revenue. When the unit ad revenue increases, this component decreases, which means that the provider should lower its price and motivate more users to travel on the arc and watch ads. The second component is affected by the effective resistances. Specifically, if a location's outgoing arcs have large unit ad revenues, it is ``valuable'' to have available vehicles at this location. The second component increases with the effective resistances between $j$ (i.e., the destination of arc $\left(i,j\right)$) and those ``valuable'' locations. This means that when sending vehicles from $i$ to $j$ makes it harder to route more vehicles to the ``valuable'' locations, the provider should increase its price on arc $\left(i,j\right)$ to reduce the traffic from $i$ to $j$. Furthermore, when the arc's travel time is long, the second component has a small impact on the optimal price. 

{\bf{2. Provider's Payoff and Consumer Surplus (Theorem \ref{theorem:payoff2CS}).}} Given the unit ad revenues, we derive the provider's payoff and consumer surplus under the provider's optimal vehicle assignment and pricing decisions.{\footnote{When analyzing the consumer surplus, we do not consider the disutility caused by watching ads. As shown in VUGO's example \cite{Vugo}, when displaying ads on tablets, a provider also offers entertainment options (e.g., watching TV shows) to users, which can compensate for the disutility of watching ads.}} First, we prove that the provider's total payoff is twice the total consumer surplus, although this proportional relation in general may not hold for the provider's payoff and consumer surplus on an arc. Second, both the provider's total payoff and the total consumer surplus are non-decreasing in an arc's unit ad revenue. If the provider receives sponsorship from an advertiser and displays its ads to the users on an arc, the provider will change its prices in the network based on the arc's unit ad revenue. Although the provider may increase some arcs' prices, which reduces the corresponding arcs' consumer surplus, we show that the total consumer surplus will not decrease in this case. 

{\bf{3. Provider's Collaboration with Advertisers (Theorems \ref{theorem:arcbad} and \ref{theorem:locationad}).}} When the provider starts implementing the in-vehicle advertising, it may only be able to collaborate with a limited number of advertisers, who have different target arcs to display ads. Different collaboration choices lead to different unit ad revenues, and we study the provider's collaboration choice that maximizes its payoff. First, we consider the situation where the provider can only select one \emph{arc-based advertiser}, who is only willing to advertise on one arc, to collaborate with. We show that the provider should collaborate with the advertiser whose target arc has a large traffic demand, a long travel time, and a small effective resistance. Intuitively, when an arc $\left(i,j\right)$ has a small effective resistance, there are many paths between $i$ and $j$. In this case, the provider can route a large mass of vehicles from other arcs to this arc, which enables the provider to increase the vehicle supply on this arc and let more users watch ads. Second, we study the provider's collaboration with \emph{location-based advertisers}, who are only willing to advertise on the incoming arcs of their target locations. When (i) a location's incoming arcs have small effective resistances and (ii) the effective resistances between these incoming arcs' origins are large, the provider can simultaneously route a large mass of vehicles to these incoming arcs. Hence, the provider should collaborate with the corresponding location-based advertiser. 

In this work, we assume that the parameters (e.g., traffic demand and unit ad revenues) are constant, and study the system's steady state (similar assumptions were made in \cite{bimpikis2016spatial,lagos2000alternative}). In practice, the parameters may change significantly across different time periods (e.g., peak hours and non-peak hours). The provider can implement different spatial pricing strategies during these periods. When selecting the advertisers to collaborate with, the provider should make the decision to maximize its average payoff over all periods. 

\subsection{Related Work}

Our work is related to the following three streams of literature.

\subsubsection{Advertising in Vehicle Networks}

There have been many papers studying advertising in vehicular ad hoc networks (VANETs), e.g., \cite{lee2007secure,qin2016post,zheng2015optimizing,zhang2017bus}. Vehicles and roadside infrastructure can propagate digital ads in VANETs via vehicle-to-vehicle communications and roadside-to-vehicle communications, respectively. Qin \emph{et al.} in \cite{qin2016post} analyzed the selection of seed vehicles, which forward ads to their neighboring vehicles, to maximize the advertising coverage. Lee \emph{et al.} in \cite{lee2007secure} proposed an ad dissemination scheme, which incentivizes selfish vehicles to forward commercial ads and prevents malicious vehicles from sending out dummy ads. Zeng and Wu in \cite{zheng2015optimizing} considered a shopkeeper who uses roadside access points to disseminate the shop's ads to the passing vehicles, and analyzed the optimal roadside access point placement. 
In our work, we consider a different advertising scenario, where a vehicle service provider displays ads in its vehicles via tablets (the ads can be preloaded or sent to the tablets via cellular network). The provider increases the mass of users watching ads through reducing its service prices rather than disseminating the ads to other vehicles. 

\subsubsection{Spatial Pricing for Vehicle Service} There is a large literature on vehicle service pricing, e.g., \cite{bimpikis2016spatial,banerjee2016pricing,banerjee2015pricing,fang2017prices,ma2018spatio}. Banerjee \emph{et al.} in \cite{banerjee2016pricing} modeled a shared vehicle system by a closed-queueing network, and developed an approximation framework for designing pricing schemes. 
Ma \emph{et al.} in \cite{ma2018spatio} proposed a spatial-temporal pricing mechanism to maximize social welfare. Bimpikis \emph{et al.} in \cite{bimpikis2016spatial} considered a ride-sharing platform, and studied the spatial pricing for riders and spatial compensation for drivers. The authors investigated a model with time-invariant traffic demand, which is similar to our work. Different from these studies, we focus on analyzing the impact of ad revenues on a provider's spatial pricing, and introduce a novel resistance-based analysis approach.

\subsubsection{Analysis of Ad-Sponsored Business Models} Some references studied other service providers who display ads to their users and receive payment from advertisers. In \cite{yu2017public}, a public Wi-Fi network provider delivers ads to its Wi-Fi users to monetize the Wi-Fi service. In \cite{guo2017economic}, an app developer allows users to get virtual items in the app by watching ads. These service providers do not need to consider flow balance constraints, which constitute a key challenge to our study.

\section{Model}\label{sec:model}
\subsection{Users' Demand}\label{subsec:userdemand}
We consider a discrete-time model, and denote the set of locations by ${\mathcal N}\triangleq \left\{1,2,\ldots,N\right\}$. In each time slot, a continuum of users of mass $\theta_{ij}\ge0$ consider taking the provider's vehicle service to \emph{depart} from location $i$ to location $j$ ($i,j\in{\mathcal N}$). In particular, we have $\theta_{ii}=0$ for all $i\in{\mathcal N}$. Note that $\left\{\theta_{ij}\right\}_{i,j\in{\mathcal N}}$ induces a directed graph ${\mathcal G}_{\rm d}=\left({\mathcal N},{\mathcal A}\right)$, where there is an arc from $i$ to $j$ (i.e., $\left(i,j\right)\in{{\mathcal A}}$) if and only if $\theta_{ij}>0$.{\footnote{In this work, we use ``an arc'' to refer to a \emph{directed} edge in ${\mathcal G}_{\rm d}$, and will use ``an edge'' to refer to an \emph{undirected} edge in an undirected graph (which will be defined later).}} We focus on the case where ${\mathcal G}_{\rm d}$ is weakly connected. When ${\mathcal G}_{\rm d}$ is disconnected, the provider can solve the problem for each weakly connected component separately. 

We use $\left\{\xi_{ij}\right\}_{i,j\in{\mathcal N}}$ to denote the travel time between different locations. If a user takes the vehicle service, the number of time slots required for the travel from location $i$ to $j$ is $\xi_{ij}>0$ ($j\ne i$). In Fig. \ref{graph:a}, we illustrate one example of ${\mathcal G}_{\rm d}$, and also indicate the users' demand (i.e., $\left\{\theta_{ij}\right\}_{i,j\in{\mathcal N}}$) and travel time (i.e., $\left\{\xi_{ij}\right\}_{i,j\in{\mathcal N}}$).

{{We use a \emph{reservation price} to refer to the highest price that a user is willing to pay.}} If the vehicle service's price is no greater than the reservation price, the user is willing to take the provider's vehicle service; otherwise, the user will travel to its destination by other approaches (e.g., taking public buses). We first assume that the reservation prices of all the $\theta_{ij}$ users on arc $\left(i,j\right)\in{\mathcal A}$ are uniformly distributed in $\left[0,1\right]$. In Section \ref{sec:numerical}, we will relax this assumption,{\footnote{Reference \cite{bimpikis2016spatial} assumed that riders' reservation prices follow a uniform distribution. Many other references also considered the consumers with uniformly distributed reservation prices \cite{kuo2011dynamic,besanko1990optimal}.}} and consider exponentially distributed reservation prices. Therefore, if the vehicle service's price is $p$, the mass of users that want to take the service from $i$ to $j$ is $\theta_{ij} \max\left\{1-p,0\right\}$. We call $\theta_{ij} \max\left\{1-p,0\right\}$ the \emph{actual demand} on arc $\left(i,j\right)$.

\subsection{Provider's Decisions, Ad Revenue, and Cost}
The provider's decisions include the vehicle assignment and pricing. We use $q_{ij}$ to denote the mass of vehicles that {depart} from $i$ to $j$ in each time slot. We use $p_{ij}$ to denote the vehicle service's price for arc $\left(i,j\right)$, measured in dollars per time slot. If a user travels from $i$ to $j$ via the vehicle service, since the travel time is $\xi_{ij}$, the user's total payment is $\xi_{ij}p_{ij}$. 

Besides the users' payment, the provider gets revenue by displaying ads to the users. In each time slot, the provider displays a fixed number of ads to each user in the provider's vehicles. To achieve targeted advertising, the ads are displayed based on the users' origins and destinations. Different origin-destination pairs can correspond to different values of ad revenue. Let $a_{ij}\ge0$ denote the provider's \emph{unit ad revenue} (i.e., the ad revenue per user per time slot) on arc $\left(i,j\right)$. If a user travels from $i$ to $j$ via the vehicle service, the provider can get $\xi_{ij}a_{ij}$ by displaying ads to the user for $\xi_{ij}$ time slots.

Next, we model the provider's cost to provide the vehicle service. When the provider is a taxi company, the cost can include the company's payment to the drivers. When the provider is an autonomous vehicle service provider, the cost can be the vehicles' energy cost. We use $c$ to denote the provider's \emph{unit cost}, and the provider's total cost for transporting a user from $i$ to $j$ is $\xi_{ij}c$. In this work, we focus on the case where $c<1$, i.e., the unit cost is smaller than the users' maximum reservation price. Furthermore, we consider a homogeneous unit cost $c$ for different arcs. 
With more cumbersome notations, we can easily extend our results to the heterogeneous unit cost case. 

\subsection{Provider's Problem}\label{subsec:providerpro}

The provider decides the vehicle assignment and pricing to maximize its time-average payoff in the system's steady state. We first assume that the provider's supply of vehicles on each arc is no greater than the actual demand, i.e., $q_{ij}\le \theta_{ij} \max\left\{1-p_{ij},0\right\}$ for all $\left(i,j\right)\in{\mathcal A}$. This implies that each vehicle sent on arc $\left(i,j\right)$ carries a user. In Section \ref{sec:numerical}, we will relax this assumption and consider the provider's strategy of routing empty vehicles in the network. Then, we can formulate the provider's problem as follows:

\begin{subequations}
\begin{align}
& \max {~} \sum_{\left(i,j\right)\in{\mathcal A}} q_{ij} \xi_{ij} \left(p_{ij}+a_{ij}-c\right)\label{opt:objective}\\
& {\rm s.t.}{~~}{~~}{~~}{~~}{~~} q_{ij}\le \theta_{ij} \max\left\{1-p_{ij},0\right\},\forall \left(i,j\right)\in{\mathcal A},\label{opt:constraint:demand}\\
& {~~}{~~}{~~}{~~}{~~}{~~}{~~}{~~}{~~}\sum_{j:\left(i,j\right)\in{\mathcal A}} q_{ij} = \sum_{j:\left(j,i\right)\in{\mathcal A}} q_{ji},\forall i\in{\mathcal N},\label{opt:constraint:balance}\\
& {\rm var.}{~~}{~~}{~~}{~~}{~~}q_{ij}\ge0,p_{ij},\forall \left(i,j\right)\in{\mathcal A}.\label{opt:var}
\end{align}\label{opt:sumsub:original}
\end{subequations}
The objective function in (\ref{opt:objective}) captures the provider's time-average payoff. In each time slot, a continuum of vehicles of mass $q_{ij}$ \emph{depart} from $i$ to $j$, i.e., $q_{ij}$ is the vehicle departure rate for the travel from $i$ to $j$. In any time slot, the mass of vehicles traveling on arc $\left(i,j\right)$ is $q_{ij} \xi_{ij}$. Each vehicle carries a user, and the provider gets $p_{ij}+a_{ij}-c$ by serving the user \emph{in this time slot}. Therefore, the provider's time-average payoff from arc $\left(i,j\right)$ in the system's steady state is $q_{ij} \xi_{ij} \left(p_{ij}+a_{ij}-c\right)$.

Constraint (\ref{opt:constraint:demand}) states that the supply of vehicles is no greater than the actual demand (which is the assumption made above). We will show that the optimal supply equals the actual demand, and hence there is no excess demand. Constraint (\ref{opt:constraint:balance}) captures the vehicle flow balance at each location $i$. For arc $\left(i,j\right)$, both the rate that the vehicles depart $i$ and the rate that the vehicles arrive at $j$ are $q_{ij}$. Considering all arcs, the vehicles' departure rate at $i$ is $\sum_{j:\left(i,j\right)\in{\mathcal A}} q_{ij}$, and the arrival rate at $i$ is $\sum_{j:\left(j,i\right)\in{\mathcal A}} q_{ji}$. The two rates are equal under constraint (\ref{opt:constraint:balance}). Note that the travel time $\xi_{ij}$ does not affect the departure rate and arrival rate, so it does not appear in the flow balance constraint. 

As indicated in (\ref{opt:var}), the provider can set negative prices. Intuitively, when the unit ad revenue is large, the provider can pay the users to motivate them to take the vehicle service. For arc $\left(i,j\right)$, $\theta_{ij}$ can be understood as the mass of users who are willing to take the vehicle service when $p_{ij}=0$. If $p_{ij}<0$, some users who originally plan to travel by other approaches will take the provider's service, and the actual demand $\theta_{ij}\left(1-p_{ij}\right)$ will be greater than $\theta_{ij}$. Under reasonable parameter settings (e.g., the settings in our numerical experiments in Section \ref{subsec:numerical:validate}), the average unit ad revenue in the network is smaller than the unit cost, and the optimal prices are usually non-negative. 

In problem (\ref{opt:sumsub:original}), we assume that the provider has sufficient vehicles to meet the actual demand. We will relax this assumption in Section \ref{sec:numerical}. Next, we transform problem (\ref{opt:sumsub:original}) to a pricing problem. 
First, we can see that we only need to focus on $p_{ij}\le1$ for all $\left(i,j\right)\in{\mathcal A}$. This is because any $p_{ij}>1$ reduces the actual demand on arc $\left(i,j\right)$ to zero, which is the same as $p_{ij}=1$. 
Second, we only need to focus on the solutions satisfying $q_{ij}= \theta_{ij} \left(1-p_{ij}\right)$ for all $\left(i,j\right)\in{\mathcal A}$. This is because when $q_{ij}< \theta_{ij} \left(1-p_{ij}\right)$, the provider can improve its payoff by increasing $p_{ij}$. Therefore, the provider can decide its prices by solving the following problem:
\begin{subequations}
\begin{align}
& \max {~} \sum_{\left(i,j\right)\in{\mathcal A}} \theta_{ij}\xi_{ij} \left(1-p_{ij}\right)  \left(p_{ij}+a_{ij}-c\right)\label{opt2:objective}\\
& {\rm s.t.}{~~}{~~}{~~}{~~}{~~} \sum_{j:\left(i,j\right)\in{\mathcal A}} \theta_{ij} \left(1-p_{ij}\right) = \sum_{j:\left(j,i\right)\in{\mathcal A}} \theta_{ji} \left(1-p_{ji}\right),\forall i\in{\mathcal N},\label{opt2:constraint:flow}\\
& {\rm var.}{~~}{~~}{~~}{~~}{~~}p_{ij}\le1,\forall \left(i,j\right)\in{\mathcal A}.\label{opt2:var}
\end{align}\label{opt:sumsub:pricing}
\end{subequations}
In Section \ref{sec:prices}, we will analyze the above problem's optimal solution. After getting the optimal price $p_{ij}^*$ for $\left(i,j\right)\in{\mathcal A}$, we can compute the optimal vehicle assignment for arc $\left(i,j\right)$ by $q_{ij}^*= \theta_{ij} \left(1-p_{ij}^*\right)$.

\section{Optimal Spatial Pricing}\label{sec:prices}
In this section, we analyze the provider's optimal prices. We will show that the directed graph ${\mathcal G}_{\rm d}$ corresponds to an \emph{electrical network}, and the provider's optimal prices heavily depend on the resistances in the \emph{electrical network}.

Let $\lambda_i$ be the dual variable corresponding to the flow balance constraint (\ref{opt2:constraint:flow}) at location $i$, and $\mu_{ij}$ be the dual variable corresponding to $p_{ij}\le1$. Since the objective function is quadratic and concave, and the constraints are affine, problem (\ref{opt:sumsub:pricing}) is convex and the KKT conditions are necessary and sufficient for optimality. 
We use $p_{ij}^*$ to denote the optimal price and $\lambda_i^*$ and $\mu_{ij}^*$ to denote the optimal dual variables. We can derive the following proposition ({\bf{we leave all the proofs in our appendix}}).

\begin{proposition}\label{proposition:KKTprice}
The provider's optimal price for arc $\left(i,j\right)\in{\mathcal A}$ is given by
\begin{align}
p_{ij}^*=\frac{1-a_{ij}+c}{2}+\frac{\lambda_i^*-\lambda_j^*}{2\xi_{ij}}-\frac{\mu_{ij}^*}{2\xi_{ij}\theta_{ij}}.\label{equ:optimalp}
\end{align}
\end{proposition}

In (\ref{equ:optimalp}), $\lambda_i^*$ and $\lambda_j^*$ correspond to the flow balance constraints at $i$ and $j$, respectively. The value of $\lambda_i^*-\lambda_j^*$ depends on the parameters (e.g., unit ad revenues) of the other arcs as well as those of arc $\left(i,j\right)$. In Sections \ref{subsec:muzero} and \ref{subsec:mupositive}, we explicitly characterize this dependency by introducing an electrical network. The results will enable us to further characterize the dependence of $p_{ij}^*$ on the network parameters.

In (\ref{equ:optimalp}), $\mu_{ij}^*$ corresponds to the constraint $p_{ij}\le1$, and hence we have $\mu_{ij}^*\ge0$. For ease of exposition, we first analyze the case where $\mu_{ij}^*=0$ for all $\left(i,j\right)\in{\mathcal A}$ in Section \ref{subsec:muzero}. Then, we extend the results to the general case (where $\mu_{ij}^*\ge0$) in Section \ref{subsec:mupositive}. 

\subsection{$\mu_{ij}^*=0$ Case}\label{subsec:muzero}
In this subsection, we analyze $\lambda_i^*-\lambda_j^*$ in (\ref{equ:optimalp}) in the case where $\mu_{ij}^*=0$ for all $\left(i,j\right)\in{\mathcal A}$. We first compute $\lambda_i^*-\lambda_j^*$ using the flow balance constraint, and then interpret the result based on an electrical network. 

\subsubsection{Value of $\lambda_i^*-\lambda_j^*$}

We substitute the expression of $p_{ij}^*$ in (\ref{equ:optimalp}) into the flow balance constraint (\ref{opt2:constraint:flow}), and get the following equation for each $i\in{\mathcal N}$:
\begin{align}
\sum_{j:\left(i,j\right)\in{\mathcal A}} \theta_{ij} \left(\frac{1+a_{ij}-c}{2}-\frac{\lambda_i^*-\lambda_j^*}{2\xi_{ij}}\right) = \sum_{j:\left(j,i\right)\in{\mathcal A}} \theta_{ji} \left(\frac{1+a_{ji}-c}{2}-\frac{\lambda_j^*-\lambda_i^*}{2\xi_{ji}}\right).\label{equ:beforematrix}
\end{align}

\begin{figure}[t]
  \centering
  \subfigure[Directed Graph ${\mathcal G}_{\rm d}=\left({\mathcal N},{\mathcal A}\right)$.]{
    \label{graph:a}
    \includegraphics[scale=0.37]{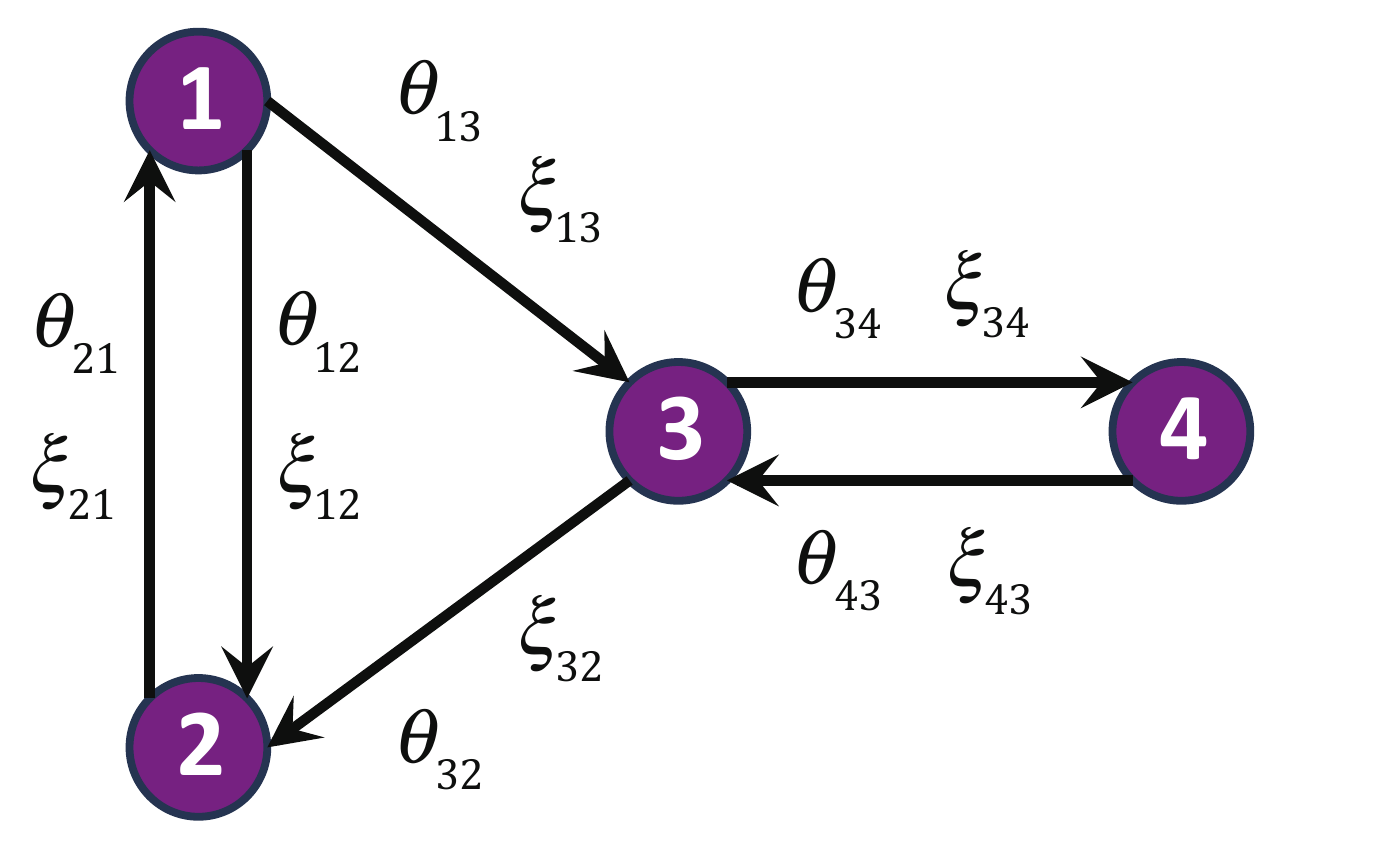}}
  \subfigure[Undirected Graph ${\mathcal G}_{\rm u}=\left({\mathcal N},{\mathcal E}\right)$.]{
    \label{graph:b}
    \includegraphics[scale=0.37]{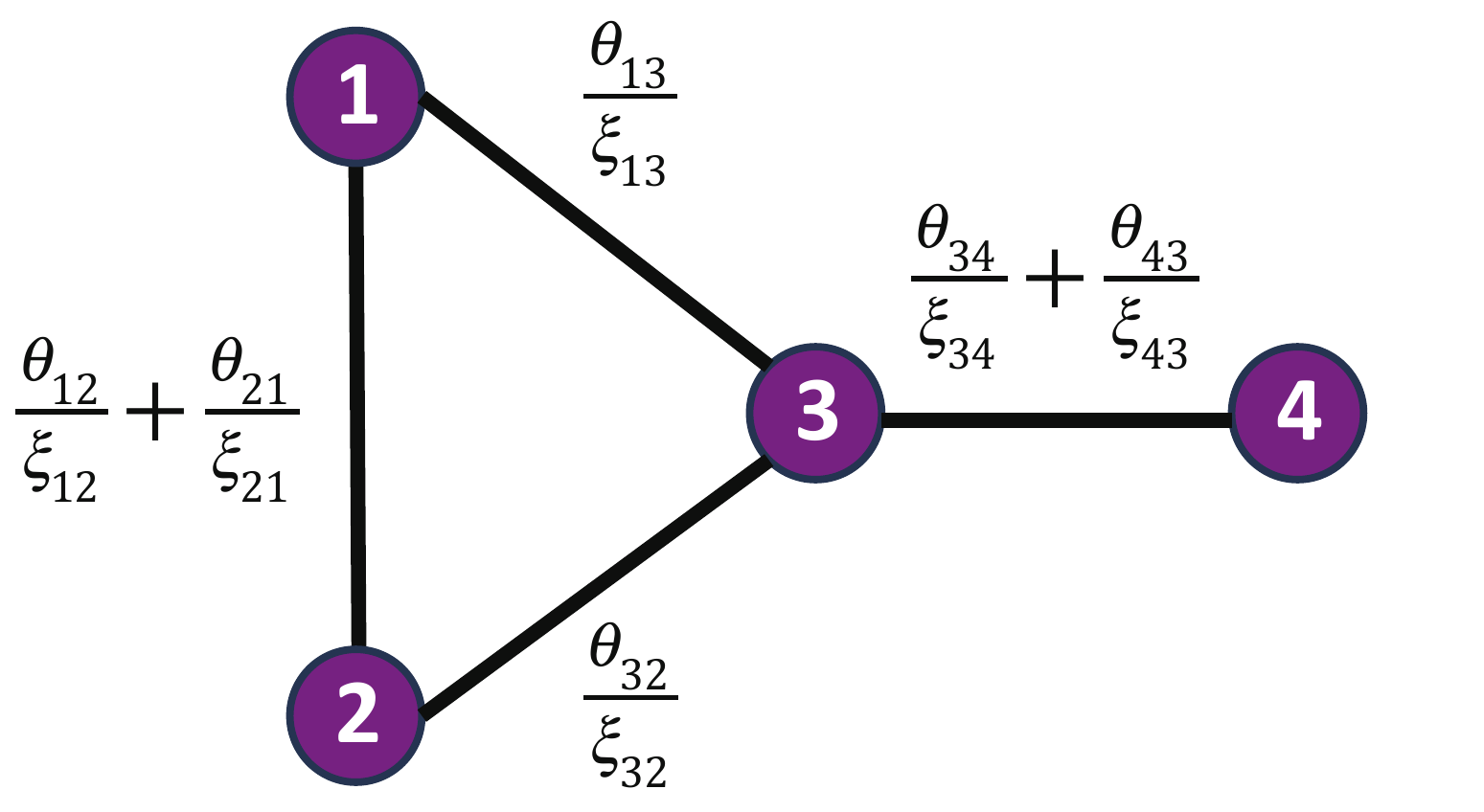}}
  \subfigure[Electrical Network.]{
    \label{graph:c}
    \includegraphics[scale=0.37]{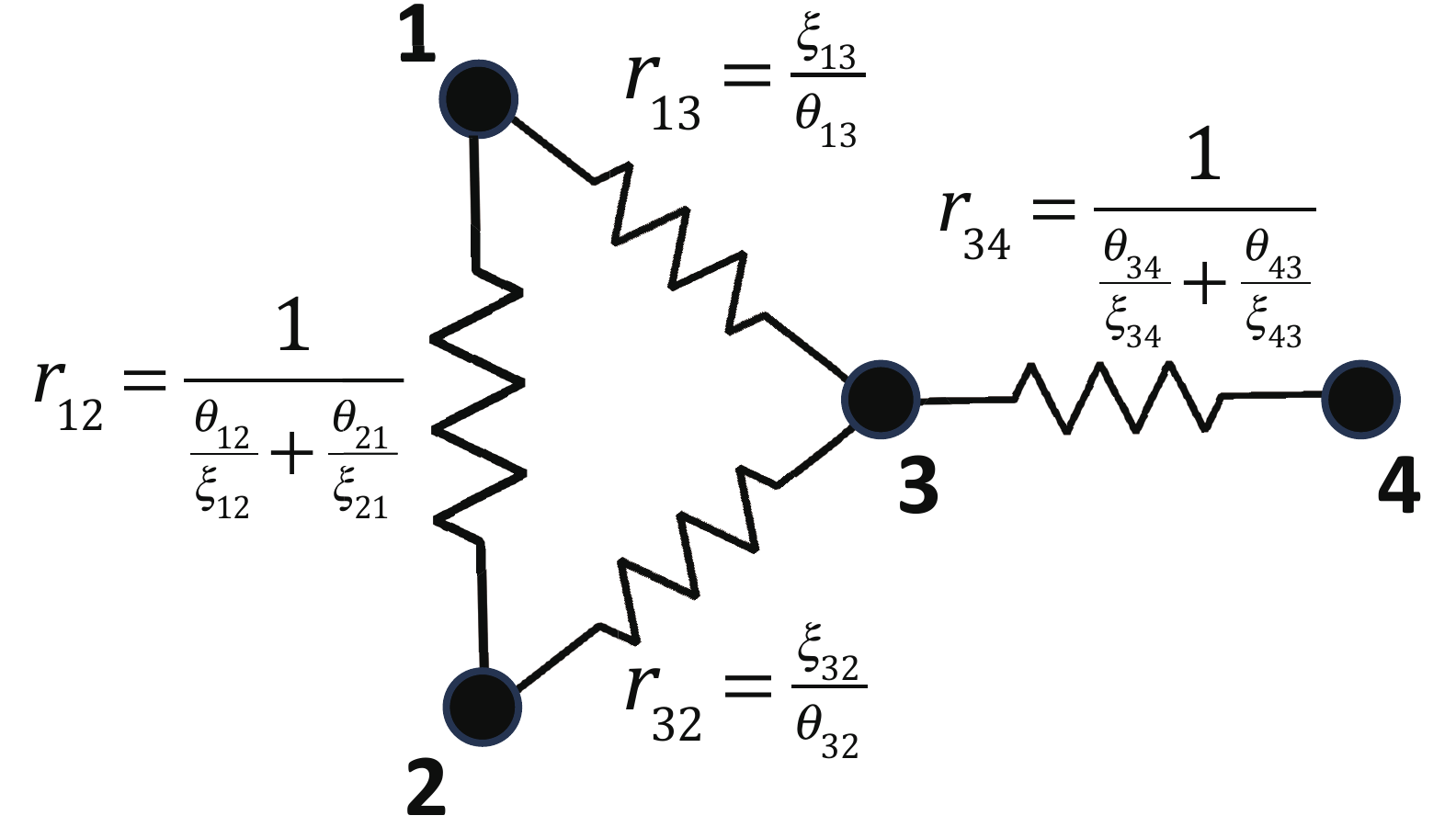}}
  \caption{Examples of ${\mathcal G}_{\rm d}$, ${\mathcal G}_{\rm u}$, and Electrical Network (When $\mu_{ij}^*=0$ for All $\left(i,j\right)\in{\mathcal A}$).}
  \label{graphabc}
\end{figure}

We can get $\lambda_i^*$ ($i\in{\mathcal N}$) by solving the system of equations given by (\ref{equ:beforematrix}). Next, we introduce several definitions, and rearrange the system of equations. First, we define an $N \times N$ matrix ${\bm L}$, whose $ij$-th entry is given by
\begin{align}
l_{ij} \triangleq \left\{ {\begin{array}{*{20}{l}}
{\sum_{m\ne i,m\in{\mathcal N}}\left(\frac{\theta_{im}}{\xi_{im}}+\frac{\theta_{mi}}{\xi_{mi}}\right),}&{{\rm if~}i=j,}\\
{-\frac{\theta_{ij}}{\xi_{ij}}-\frac{\theta_{ji}}{\xi_{ji}},}&{{\rm if~}i\ne j.}\\
\end{array}} \right.
\end{align}
Matrix ${\bm L}$ is the Laplacian matrix of a weighted undirected graph ${\mathcal G}_{\rm u}=\left({\mathcal N},{\mathcal E}\right)$, which corresponds to the directed graph ${\mathcal G}_{\rm d}$. Specifically, ${\mathcal E}$ contains edge $\left(i,j\right)$ if and only if there exists at least one arc between $i$ and $j$ in the directed graph ${\mathcal G}_{\rm d}$. Moreover, edge $\left(i,j\right)$ is associated with a weight, which is $\frac{\theta_{ij}}{\xi_{ij}}+\frac{\theta_{ji}}{\xi_{ji}}$. We illustrate an example of ${\mathcal G}_{\rm u}$ in Fig. \ref{graph:b}. Since ${\mathcal G}_{\rm d}$ is weakly connected, ${\mathcal G}_{\rm u}$ is connected.

Second, we define an $N \times 1$ vector $\bm v$, whose $i$-th entry is given by
\begin{align}
v_i \triangleq \sum_{m:\left(i,m\right)\in{\mathcal A}} \theta_{im} \left(1+a_{im}-c\right)- \sum_{m:\left(m,i\right)\in{\mathcal A}} \theta_{mi} \left(1+a_{mi}-c\right).\label{equ:valueoflocation}
\end{align}
Here, $v_i$ measures the value of having available vehicles at location $i$. When $v_i$ is large, the traffic demand and unit ad revenues on the arcs originating from $i$ are large (compared with those on the arcs pointing to $i$). In this situation, the provider needs to route more vehicles to $i$ to achieve a high payoff. Note that $v_i$ can be negative, and we always have $\sum_{i\in{\mathcal N}} v_i=0$.

Let ${\bm \lambda}^*\triangleq \left(\lambda_1^*,\lambda_2^*,\ldots, \lambda_N^*\right)^\intercal$. Using the fact that $\theta_{ij}=0$ for all $\left(i,j\right)\notin {\mathcal A}$, we can rearrange the system of equations given by (\ref{equ:beforematrix}) as
\begin{align}
{\bm L} {\bm \lambda}^*={\bm v}.\label{equ:linearsystem}
\end{align}
Note that ${\bm L}$ is the Laplacian matrix of a connected undirected graph ${\mathcal G}_{\rm u}$. Hence, the rank of ${\bm L}$ is $N-1$, and ${\bm L}$ is non-invertible \cite{ranjan2014incremental}. As a substitute for the inverse, we consider the generalized inverse of ${\bm L}$ \cite{dorfler2018electrical}, and denote it by ${\bm L}^+$. In the following proposition, we characterize the solution space of (\ref{equ:linearsystem}) based on ${\bm L}^+$.

\begin{proposition}\label{proposition:solutionspace}
The solution space of (\ref{equ:linearsystem}) is
\begin{align}
\left\{{\bm \lambda}: {\bm \lambda}={\bm L}^+ {\bm v} + \beta \left(1,1,\ldots,1\right)^\intercal, \beta\in{\mathbb R}\right\}.
\end{align}
Moreover, we have
\begin{align}
\lambda_i^*-\lambda_j^*=\sum_{k\in{\mathcal N}} \left(l_{ik}^+ - l_{jk}^+\right) v_k,\forall i,j\in{\mathcal N}.\label{equ:gaplambda}
\end{align}
\end{proposition}
There are infinitely many ${\bm \lambda}^*$ satisfying (\ref{equ:linearsystem}), but (\ref{equ:gaplambda}) shows that $\lambda_i^*-\lambda_j^*$ is uniquely determined by $l_{ik}^+$, $l_{jk}^+$, and $v_k$, $k\in{\mathcal N}$. Next, we interpret $l_{ik}^+$ and $l_{jk}^+$ by introducing an electrical network. 

\subsubsection{Electrical Network}
Given graph ${\mathcal G}_{\rm u}=\left({\mathcal N},{\mathcal E}\right)$, we can get an electrical network via replacing the vertices in ${\mathcal N}$ by nodes and the edges in ${\mathcal E}$ by resistors. If $\left(i,j\right)\in {\mathcal E}$, we use $r_{ij}$ to denote the resistance of the resistor between nodes $i$ and $j$, and let $r_{ij}=\frac{1}{\frac{\theta_{ij}}{\xi_{ij}}+\frac{\theta_{ji}}{\xi_{ji}}}$ (which is the inverse of the weight of edge $\left(i,j\right)$ in ${\mathcal G}_{\rm u}$). In Fig. \ref{graph:c}, we illustrate an example of the electrical network.

In the electrical network, the \emph{effective resistance} between any two nodes $i$ and $j$ ($i,j\in{\mathcal N}$) is defined as the voltage between $i$ and $j$ when a unit current is injected at $i$ and withdrawn at $j$. Let $R_{ij}$ denote the effective resistance between $i$ and $j$. We can see that $R_{ii}=0$ and $R_{ij}=R_{ji}$ for all $i,j\in{\mathcal N}$. For example, in Fig. \ref{graph:c}, $R_{12}$ can be computed as $R_{12}=\frac{r_{12}\left(r_{13}+r_{32}\right)}{r_{12}+r_{13}+r_{32}}$.

According to \cite{ranjan2014incremental}, the effective resistances in the electrical network have the following relation with ${\bm L}^+$ of graph ${\mathcal G}_{\rm u}$:
\begin{align}
R_{ij}=l_{ii}^+ + l_{jj}^+ - 2l_{ij}^+,\forall i,j\in{\mathcal N}.\label{equ:RandLplus}
\end{align}
Using this relation, the property that $\sum_{k\in{\mathcal N}} v_k=0$, and Proposition \ref{proposition:solutionspace}, we can prove the following theorem, which characterizes $\lambda_i^*-\lambda_j^*$ and $p_{ij}^*$ based on the effective resistances.
\begin{theorem}\label{theorem:priceresistance}
In the case where $\mu_{ij}^*=0$ for all $\left(i,j\right)\in{\mathcal A}$, the value of $\lambda_i^*-\lambda_j^*$ is given by
\begin{align}
\lambda_i^*-\lambda_j^* =\frac{1}{2}\sum_{k\in{\mathcal N}} \left(R_{jk}-R_{ik}\right)v_k,\forall i,j\in{\mathcal N},\label{equ:difflambda}
\end{align}
where $v_k$ is defined in (\ref{equ:valueoflocation}). From (\ref{equ:optimalp}), we can compute $p_{ij}^*$ by
\begin{align}
p_{ij}^*=\frac{1-a_{ij}+c}{2}+\frac{1}{4\xi_{ij}} \sum_{k\in{\mathcal N}} \left(R_{jk}-R_{ik}\right) v_k,\forall \left(i,j\right)\in{\mathcal A}.\label{equ:priceresistance}
\end{align}
\end{theorem}
{\bf{Remark 1}:} Theorem \ref{theorem:priceresistance} enables us to understand the dependence of $p_{ij}^*$ on the network topology and parameters: 
\begin{itemize}
\item First, the term $\frac{1-a_{ij}+c}{2}$ is only related to arc $\left(i,j\right)$. When $a_{ij}$ increases, this term decreases, which means that the provider should lower its price and motivate more users to travel on arc $\left(i,j\right)$ to get the ad revenue. 
\item Second, $\sum_{k\in{\mathcal N}} \left(R_{jk}-R_{ik}\right) v_k$ is affected by the effective resistances and vector ${\bm v}$. The effective resistance between two nodes measures their ``distance'' in the electrical network (a large effective resistance implies a large distance). Moreover, $v_k$ measures the value of having vehicles at $k$ (as defined in (\ref{equ:valueoflocation})). Suppose that, compared with location $i$, location $j$ is ``farther'' from most of the locations with large $v_k$ in the electrical network. In this situation, $\sum_{k\in{\mathcal N}} \left(R_{jk}-R_{ik}\right) v_k$ is large. Eq. (\ref{equ:priceresistance}) implies that the provider should charge a high price $p_{ij}^*$ to reduce the actual demand on arc $\left(i,j\right)$. This reduces the mass of vehicles traveling from $i$ to $j$, which is farther from most ``valuable'' locations. 
\item Third, when $\xi_{ij}$ is large, $\sum_{k\in{\mathcal N}} \left(R_{jk}-R_{ik}\right) v_k$ has a small impact on $p_{ij}^*$. Therefore, when an arc has a long travel time, the network topology and the other arcs' parameters have a small impact on the arc's optimal price.
\end{itemize}

An interesting observation is that the direction of user demand affects $p_{ij}^*$ via vector ${\bm v}$ rather than the effective resistances. Specifically, the resistance of the resistor between two locations $x$ and $y$ is computed by $r_{xy}=\frac{1}{\frac{\theta_{xy}}{\xi_{xy}}+\frac{\theta_{yx}}{\xi_{yx}}}$. Suppose that $\xi_{xy}=\xi_{yx}$ and $\theta_{xy}\ne\theta_{yx}$. If we swap the values of $\theta_{xy}$ and $\theta_{yx}$, the value of $r_{xy}$ will not change, which implies that the effective resistances between different locations will not change. However, the swapping of $\theta_{xy}$ and $\theta_{yx}$ will change $v_x$ and $v_y$, and further affect $p_{ij}^*$. 

\subsection{General Case}\label{subsec:mupositive}
In this subsection, we discuss the extension of our results in Section \ref{subsec:muzero} to the general case, where $\mu_{ij}^*\ge0$ for all $\left(i,j\right)\in{\mathcal A}$. We will characterize $\lambda_i^*-\lambda_j^*$ and $p_{ij}^*$ based on a modified electrical network. 

We can compute $\mu_{ij}^*$ ($\left(i,j\right)\in{\mathcal A}$) based on the KKT conditions. If $\mu_{ij}^*=0$, we can simplify (\ref{equ:optimalp}) as $p_{ij}^*=\frac{1-a_{ij}+c}{2}+\frac{\lambda_i^*-\lambda_j^*}{2\xi_{ij}}$; otherwise, we have $p_{ij}^*=1$ based on the complementary slackness condition. By substituting the expression of $p_{ij}^*$ into the flow balance constraint (\ref{opt2:constraint:flow}), we get the following equations: 
\begin{align}
\sum_{j:\left(i,j\right)\in{\mathcal A}} \theta_{ij} {\mathbbm 1}_{\left\{\mu_{ij}^*=0\right\}} \left(\frac{1+a_{ij}-c}{2}-\frac{\lambda_i^*-\lambda_j^*}{2\xi_{ij}}\right) = 
\sum_{j:\left(j,i\right)\in{\mathcal A}} \theta_{ji} {\mathbbm 1}_{\left\{\mu_{ji}^*=0\right\}} \left(\frac{1+a_{ji}-c}{2}-\frac{\lambda_j^*-\lambda_i^*}{2\xi_{ji}}\right),\forall i\in{\mathcal N}.\label{equ:indicatormatrix}
\end{align}
Compared with (\ref{equ:beforematrix}) in Section \ref{subsec:muzero}, Eq. (\ref{equ:indicatormatrix}) includes an extra indicator function ${\mathbbm 1}_{\left\{\cdot\right\}}$, which equals $1$ if the event in braces is true and equals $0$ otherwise.

Let ${\hat \theta_{ij}}\triangleq \theta_{ij} {\mathbbm 1}_{\left\{\mu_{ij}^*=0\right\}} $ for all $i,j\in{\mathcal N}$. After replacing $\left\{{\theta_{ij}}\right\}_{i,j\in{\mathcal N}}$ by $\left\{{\hat \theta_{ij}}\right\}_{i,j\in{\mathcal N}}$, we can conduct a similar analysis as in Section \ref{subsec:muzero}. Specifically, we define ${\hat {\bm v}}$ and a new electrical network based on $\left\{{\hat \theta}_{ij}\right\}_{i,j\in{\mathcal N}}$, and compute the corresponding effective resistance ${\hat R}_{ij}$ for all $i,j\in{\mathcal N}$. Note that when ${\hat \theta_{ij}}={\hat \theta_{ji}}=0$, nodes $i$ and $j$ are not directly connected by a resistor in the new electrical network. Similar to (\ref{equ:difflambda}) in Theorem \ref{theorem:priceresistance}, the value of $\lambda_i^*-\lambda_j^*$ is $\frac{1}{2}\sum_{k\in{\mathcal N}} \left({\hat R}_{jk}-{\hat R}_{ik}\right){\hat v}_k$. Therefore, in the general case, the optimal price $p_{ij}^*$ ($\left(i,j\right)\in{\mathcal A}$) satisfies the following equation:
\begin{align}
p_{ij}^*= \left\{ {\begin{array}{*{20}{l}}
{\frac{1-a_{ij}+c}{2}+\frac{1}{4\xi_{ij}} \sum_{k\in{\mathcal N}} \left({\hat R}_{jk}-{\hat R}_{ik}\right) {\hat v}_k,}&{{\rm if~}\mu_{ij}^*=0,}\\
{1,}&{{\rm if~}\mu_{ij}^*>0.}\\
\end{array}} \right.\label{equ:general:optimalprice}
\end{align}
Eq. (\ref{equ:general:optimalprice}) reduces to (\ref{equ:priceresistance}) in Theorem \ref{theorem:priceresistance} when $\mu_{ij}^*=0$ for all $\left(i,j\right)\in{\mathcal A}$. Based on (\ref{equ:general:optimalprice}), we can see the connection between the optimal prices and an electrical network, and derive similar insights as in {Remark 1}. 

\subsection{Discussion of $\mu_{ij}^*=0$ Case and General Case}
According to our analysis in Sections \ref{subsec:muzero} and \ref{subsec:mupositive}, the provider's optimal prices can always be characterized by the effective resistances in an electrical network. When $\mu_{ij}^*=0$ for all $\left(i,j\right)\in{\mathcal A}$, the electrical network is defined based on $\left\{{\theta_{ij}}\right\}_{i,j\in{\mathcal N}}$; otherwise, it is defined based on $\left\{{\hat \theta_{ij}}\right\}_{i,j\in{\mathcal N}}$. In Sections \ref{sec:arcs}-\ref{sec:guide}, we will first derive the results for the $\mu_{ij}^*=0$ case, which simplifies the presentation, and then extend the results to the general case. 

To understand the characteristics of the network parameters that lead to $\mu_{ij}^*=0$, we introduce Proposition \ref{proposition:asymbound}. It shows a sufficient condition under which we have $\mu_{ij}^*=0$ for all $\left(i,j\right)\in{\mathcal A}$.

\begin{proposition}\label{proposition:asymbound}
When the system parameters satisfy
\begin{align}
\sum_{k\in{\mathcal N}} \left|v_k\right| \le \min_{\left(x,y\right)\in{\mathcal A}} 2\left(\theta_{xy}+\theta_{yx}\frac{\xi_{xy}}{\xi_{yx}}\right)\left(1+a_{xy}-c\right),\label{equ:asymbound}
\end{align}
we have $\mu_{ij}^*=0$ for all $\left(i,j\right)\in{\mathcal A}$.
\end{proposition}

Intuitively, (\ref{equ:asymbound}) means that the arcs originating from and pointing to each location have similar traffic demand and unit ad revenues. Specifically, $v_k$ is defined in (\ref{equ:valueoflocation}), and equals the difference between $\sum_{m:\left(k,m\right)\in{\mathcal A}} \theta_{km} \left(1+a_{km}-c\right)$ and $\sum_{m:\left(m,k\right)\in{\mathcal A}} \theta_{mk} \left(1+a_{mk}-c\right)$. The first term corresponds to the arcs originating from $k$, and the second term corresponds to the arcs pointing to $k$. Hence, $\sum_{k\in{\mathcal N}} \left|v_k\right|$ captures the sum of the absolute differences between these two terms over all locations. The right side of (\ref{equ:asymbound}) is a bound determined by the network parameters. 

For example, if $\theta_{ij}=\theta_{ji}$ and $a_{ij}=a_{ji}$ for all $\left(i,j\right)\in{\mathcal A}$, we can verify that $v_k=0$ for all $k\in{\mathcal N}$, and inequality (\ref{equ:asymbound}) always holds (recall that $c<1$). Moreover, if ${\mathcal G}_{\rm d}$ corresponds to a unidirectional ring network and the arcs have the same traffic demand and unit ad revenue, we also have $v_k=0$ for all $k\in{\mathcal N}$, and (\ref{equ:asymbound}) is satisfied. 

\section{Influence Between Arcs}\label{sec:arcs}
In this section, we study the impact of an arc's unit ad revenue on the provider's optimal prices for different arcs. We first focus on the $\mu_{ij}^*=0$ case (e.g., when the system parameters satisfy (\ref{equ:asymbound}) in Proposition \ref{proposition:asymbound}), and introduce the following theorem. 

\begin{theorem}\label{theorem:adarc}
For any $\left(x,y\right)\in{\mathcal A}$, its unit ad revenue's impact on the optimal price for arc $\left(i,j\right)\in{\mathcal A}$ is given by
\begin{align}
\frac{\partial p_{ij}^*}{\partial a_{xy}}=-\frac{1}{2}{\mathbbm 1}_{\left\{i=x,j=y\right\}}+\frac{\theta_{xy}}{4\xi_{ij}}\left(R_{jx}-R_{ix}-R_{jy}+R_{iy}\right).\label{equ:impactadprice}
\end{align}
In particular, we have $\frac{\partial p_{xy}^*}{\partial a_{xy}}\le0$.
\end{theorem}
{\bf{Remark 2}:} Theorem \ref{theorem:adarc} characterizes the impact of $a_{xy}$ on $p_{ij}^*$: 
\begin{itemize}
\item First, $p_{xy}^*$ is non-increasing in $a_{xy}$, i.e., $\frac{\partial p_{xy}^*}{\partial a_{xy}}\le0$. When $a_{xy}$ increases, the provider should reduce $p_{xy}$, which motivates more users to travel on arc $\left(x,y\right)$ and generates more ad revenue. 
\item Second, if $\left(i,j\right)\ne\left(x,y\right)$, the sign of $\frac{\partial p_{ij}^*}{\partial a_{xy}}$ is determined by the sign of $R_{jx}-R_{ix}-\left(R_{jy}-R_{iy}\right)$. 
Specifically, when $a_{xy}$ increases, the provider needs more vehicles to travel on arc $\left(x,y\right)$ to serve more users on this arc. The flow balance constraints for locations $x$ and $y$ require the provider to route more vehicles to $x$ and less vehicles to $y$, respectively. 
Next, we show that if $R_{jx}-R_{ix}-\left(R_{jy}-R_{iy}\right)<0$, decreasing $p_{ij}$ can route more vehicles to be \emph{closer to $x$} or \emph{farther from $y$}.{\footnote{In this work, we use ``closer'' and ``farther'' to describe the change in the distance measured by the effective resistance in the electrical network rather than the physical distance. The effective resistance is affected by the traffic demand and travel time. Hence, when two locations are close to each other in the electrical network, there can be a large physical distance between them.}} 
If $R_{jx}-R_{ix}-\left(R_{jy}-R_{iy}\right)<0$, at least one of $R_{jx}-R_{ix}<0$ and $R_{jy}-R_{iy}>0$ holds. 
Recall that the effective resistance between two locations measures their distance in the electrical network. 
The inequality $R_{jx}-R_{ix}<0$ means that location $x$ is closer to $j$ than to $i$, and $R_{jy}-R_{iy}>0$ means that location $y$ is farther from $j$ than from $i$. 
Therefore, if $R_{jx}-R_{ix}-\left(R_{jy}-R_{iy}\right)<0$, a vehicle can be \emph{closer to $x$} or \emph{farther from $y$} by traveling from $i$ to $j$. 
When the provider decreases $p_{ij}$, the actual demand on arc $\left(i,j\right)$ increases and more vehicles travel from $i$ to $j$. In this case, more vehicles are routed to be closer to $x$ or farther from $y$. 
The above discussion shows that $R_{jx}-R_{ix}-\left(R_{jy}-R_{iy}\right)<0$ leads to $\frac{\partial p_{ij}^*}{\partial a_{xy}}<0$, which is consistent with (\ref{equ:impactadprice}). 
\item Third, Theorem \ref{theorem:adarc} implies that if $\theta_{xy}$ (i.e., the traffic demand on $\left(x,y\right)$) is large and $\xi_{ij}$ (i.e., the travel time on $\left(i,j\right)$) is small, $a_{xy}$ has a large impact on $p_{ij}^*$.
\end{itemize}

{{It is important to use the resistance-based analysis to derive the impact of $a_{xy}$ on $p_{ij}^*$. For example, if we do not interpret $\lambda_i^*-\lambda_j^*$ by effective resistances, it will be challenging to analyze the impact of $a_{xy}$ on $\lambda_i^*-\lambda_j^*$ in (\ref{equ:optimalp}).}}

Based on Theorem \ref{theorem:adarc}, we can analyze the sign of $\frac{\partial p_{ij}^*}{\partial a_{xy}}$ when arcs $\left(i,j\right)$ and $\left(x,y\right)$ have one common vertex. We introduce the following corollary.
\begin{corollary}\label{corollary:comnode}
For any $\left(x,y\right)\in{\mathcal A}$, its unit ad revenue's impact on the optimal price for arc $\left(x,j\right)\in{\mathcal A}$ ($j\ne y$) is given by $\frac{\partial p_{xj}^*}{\partial a_{xy}}=\frac{\theta_{xy}}{4\xi_{xj}}\left(R_{jx}-R_{jy}+R_{xy}\right)$, which is non-negative. 
\end{corollary}
The effective resistances among three locations satisfy the triangle inequality \cite{dorfler2018electrical}, which leads to $R_{jx}-R_{jy}+R_{xy}\ge0$. Corollary \ref{corollary:comnode} shows that when $a_{xy}$ increases, the provider can increase $p_{xj}$ ($j\ne y$). This reduces the actual demand on $\left(x,j\right)$, and enables the provider to offer more vehicles on $\left(x,y\right)$. 
Similar to Corollary \ref{corollary:comnode}, we can analyze $\frac{\partial p_{yj}^*}{\partial a_{xy}}$, $\frac{\partial p_{ix}^*}{\partial a_{xy}}$, and $\frac{\partial p_{iy}^*}{\partial a_{xy}}$ ($j\ne y, i\ne x,y$).

Next, we analyze some specific network topologies, and show that an arc's price can be independent of the unit ad revenue on another arc. Recall that the traffic demand (i.e., $\left\{\theta_{ij}\right\}_{i,j\in{\mathcal N}}$) and travel time (i.e., $\left\{\xi_{ij}\right\}_{i,j\in{\mathcal N}}$) define the edge weights of an undirected graph ${\mathcal G}_{\rm u}$. First, we introduce the following Proposition.

\begin{proposition}\label{proposition:completegraph}
Suppose that ${\mathcal G}_{\rm u}$ is a complete graph and all edges have the same weight. If arcs $\left(i,j\right)$ and $\left(x,y\right)$ do not have a common vertex, $p_{ij}^*$ does not change with $a_{xy}$.
\end{proposition}

When ${\mathcal G}_{\rm u}$ is a complete graph and has homogeneous edge weights, the effective resistances between different locations are the same. In this case, a vehicle's distance to $x$ or $y$ does not change after it travels from $i$ to $j$ ($i\ne x,y$ and $j\ne x,y$). Based on our discussion in {Remark 2}, this implies $\frac{\partial p_{ij}^*}{\partial a_{xy}}=0$. Note that Proposition \ref{proposition:completegraph} only requires $\frac{\theta_{ij}}{\xi_{ij}}+\frac{\theta_{ji}}{\xi_{ji}}$ (i.e., the edge weight) to be the same for all $i\ne j$. The values of $\theta_{ij}$ and $\theta_{ji}$ (\emph{or} $\xi_{ij}$ and $\xi_{ji}$) can be different. 

Second, we consider the situation where ${\mathcal G}_{\rm u}$ contains cut-vertices. When $k$ is a cut-vertex, we can remove it and its associated edges from ${\mathcal G}_{\rm u}$ to get a graph (denoted by ${\mathcal G}_{\rm u}-k$) that has more than one connected component. We introduce the following proposition.

\begin{proposition}\label{proposition:cutvertex}
Suppose that $k\in{\mathcal N}$ is a cut-vertex of ${\mathcal G}_{\rm u}$, and ${\mathcal N}_1$ and ${\mathcal N}_2$ are the vertex sets of two connected components of ${\mathcal G}_{\rm u}-k$. If $i,j\in{\mathcal N}_1 \cup \left\{k\right\}$ and $x,y\in{\mathcal N}_2 \cup \left\{k\right\}$, $p_{ij}^*$ does not change with $a_{xy}$.
\end{proposition}

Proposition \ref{proposition:cutvertex} implies that if the provider displays ads to the users traveling from $x$ to $y$ ($x,y\in{\mathcal N}_2 \cup \left\{k\right\}$), only the prices for the arcs that are associated with vertices in ${\mathcal N}_2 \cup \left\{k\right\}$ may change. 
We can see that this result can be applied to analyze the influence between arcs when ${\mathcal G}_{\rm u}$ is a star or a tree (e.g., the center vertex in a star graph is a cut-vertex). 

In the last part of this section, we discuss the extension of the results to the general case (where $\mu_{ij}^*\ge0$). Given the values of system parameters, we can compute $\mu_{ij}^*$ for all $\left(i,j\right)\in{\mathcal A}$ and get a modified electrical network (as discussed in Section \ref{subsec:mupositive}). 
If $\mu_{ij}^*>0$, $p_{ij}^*$ equals $1$ and does not change with $a_{xy}$; otherwise, we can characterize $\frac{\partial p_{ij}^*}{\partial a_{xy}}$ by ${\hat \theta}_{xy}$, ${\hat R}_{jx}$, ${\hat R}_{ix}$, ${\hat R}_{jy}$, and ${\hat R}_{iy}$ (the effective resistances are defined based on the modified electrical network) in a similar way as in Theorem \ref{theorem:adarc}. 
Note that as $a_{xy}$ increases from zero to a large value, the set $\left\{\left(i,j\right)\in{\mathcal A}:\mu_{ij}^*>0\right\}$ may change. When this set changes, we need to update $\left\{{\hat \theta_{ij}}\right\}_{i,j\in{\mathcal N}}$, the electrical network, and the corresponding effective resistances (i.e., ${\hat R}_{jx}$, ${\hat R}_{ix}$, ${\hat R}_{jy}$, and ${\hat R}_{iy}$). 

\section{Provider Payoff and Consumer Surplus}\label{sec:cs}
In this section, we study the provider's payoff and consumer surplus under the optimal spatial pricing, and show that they are non-decreasing in the unit ad revenues. We first focus on the $\mu_{ij}^*=0$ case. By substituting the expression of $p_{ij}^*$ in (\ref{equ:priceresistance}) into the objective function in (\ref{opt2:objective}), we can compute the provider's optimal (time-average) payoff, which is denoted by $\Pi^{\rm provider}$ and shown in the following proposition.

\begin{proposition}\label{proposition:platpayoff}
The provider's payoff under $\left\{p_{ij}^*\right\}_{\left(i,j\right)\in {\mathcal A}}$ is given by
\begin{align}
\Pi^{\rm provider}= \sum_{\left(i,j\right)\in {\mathcal A}} \theta_{ij}\xi_{ij} \left(\left(\frac{1+a_{ij}-c}{2}\right)^2 - \left(\frac{\sum_{k\in{\mathcal N}} \left(R_{jk}-R_{ik}\right) v_k}{4 \xi_{ij}}\right)^2 \right).\label{equ:opprice:payoff}
\end{align}
\end{proposition}

Then, we compute the consumer surplus under the optimal prices. In each time slot, the mass of users that \emph{depart} from $i$ to $j$ is $\theta_{ij} \left(1-p_{ij}^*\right)$, and hence the mass of users traveling on arc $\left(i,j\right)$ is $\theta_{ij} \xi_{ij} \left(1-p_{ij}^*\right)$. For each of these users, its surplus is uniformly distributed in $\left[0,1-p_{ij}^*\right]$. Therefore, the consumer surplus for arc $\left(i,j\right)$ is $\frac{1}{2} \theta_{ij} \xi_{ij} \left(1-p_{ij}^*\right)^2$. Considering all arcs in $\mathcal A$, we can compute the (time-average) total consumer surplus in the following proposition. 

\begin{proposition}\label{proposition:optimalCS}
The total consumer surplus under $\left\{p_{ij}^*\right\}_{\left(i,j\right)\in {\mathcal A}}$ is given by
\begin{align}
{\rm CS}=\sum_{\left(i,j\right)\in{\mathcal A}} \frac{1}{2} \theta_{ij} \xi_{ij} \left(\frac{1+a_{ij}-c}{2}-\frac{\sum_{k\in{\mathcal N}} \left(R_{jk}-R_{ik}\right) v_k}{4\xi_{ij} }\right)^2.\label{equ:opprice:cs}
\end{align}
\end{proposition}

It is challenging to compare $\Pi^{\rm provider}$ and ${\rm CS}$. One analysis approach is to rearrange their expressions in (\ref{equ:opprice:payoff}) and (\ref{equ:opprice:cs}) based on the properties of effective resistances and vector ${\bm v}$. For example, we need to use the following local sum rules of the resistances \cite{chen2010random}:
\begin{align}
\sum_{j:\left(i,j\right)\in{\mathcal E}} \frac{R_{ij}+R_{ik}-R_{jk}}{r_{ij}} =2, \forall i\ne k, i,k\in{\mathcal N}.\label{equ:localsum}
\end{align}
Recall that $r_{ij}$ is the resistance of the resistor between $i$ and $j$ and equals $\frac{1}{\frac{\theta_{ij}}{\xi_{ij}}+\frac{\theta_{ji}}{\xi_{ji}}}$. After rearranging (\ref{equ:opprice:payoff}) and (\ref{equ:opprice:cs}), we can prove that $\Pi^{\rm provider}$ is proportional to ${\rm CS}$, and introduce the following theorem.

\begin{theorem}\label{theorem:payoff2CS}
Under the optimal prices, the provider's payoff is twice the consumer surplus, i.e., $\Pi^{\rm provider}=2{\rm CS}$. Moreover, both $\Pi^{\rm provider}$ and ${\rm CS}$ are non-decreasing in $a_{xy}$ for any $\left(x,y\right)\in{\mathcal A}$.
\end{theorem}

{\bf{Remark 3}:} Theorem \ref{theorem:payoff2CS} shows the following results:
\begin{itemize}
\item First, $\Pi^{\rm provider}=2{\rm CS}$. Under the optimal prices, the provider's payoff from arc $\left(i,j\right)$ is $\theta_{ij}\xi_{ij} \left(1-p_{ij}^*\right)  \left(p_{ij}^*+a_{ij}-c\right)$, and the consumer surplus on this arc is $\frac{1}{2} \theta_{ij} \xi_{ij} \left(1-p_{ij}^*\right)^2$. We can compare these two expressions by substituting the result of $p_{ij}^*$ in (\ref{equ:optimalp}) into them. We can see that because of the flow balance constraints (i.e., the existence of $\lambda_i^*-\lambda_j^*$ in (\ref{equ:optimalp})), the provider's payoff from $\left(i,j\right)$ in general may not be twice the consumer surplus on this arc. However, Theorem \ref{theorem:payoff2CS} shows that under the optimal prices, the provider's total payoff is always twice the total consumer surplus. 
\item Second, both $\Pi^{\rm provider}$ and ${\rm CS}$ are non-decreasing in $a_{xy}$. It is easy to prove that $\Pi^{\rm provider}$ is non-decreasing in $a_{xy}$, because the provider's payoff under any given pricing solution is non-decreasing in $a_{xy}$. Since ${\rm CS}=\frac{1}{2}\Pi^{\rm provider}$, ${\rm CS}$ is also non-decreasing in $a_{xy}$. This means that if the provider receives sponsorship from an advertiser, displays ads to the users on $\left(x,y\right)$, and changes its prices in the network based on $a_{xy}$, the total consumer surplus will not decrease. 
However, the consumer surplus on an arc may decrease. Since the consumer surplus on arc $\left(i,j\right)$ is $\frac{1}{2} \theta_{ij} \xi_{ij} \left(1-p_{ij}^*\right)^2$, whether it decreases with $a_{xy}$ depends on $\frac{\partial p_{ij}^*}{\partial a_{xy}}$, which is characterized in Theorem \ref{theorem:adarc}. 
\end{itemize}

In the above analysis, we consider the case where the system parameters lead to $\mu_{ij}^*=0$ for all $\left(i,j\right)\in{\mathcal A}$. Next, we discuss the general case ($\mu_{ij}^*\ge0$). If $\mu_{ij}^*>0$, then $p_{ij}^*=1$ and both the provider's payoff and consumer surplus on $\left(i,j\right)$ are zero. We can compute $\Pi^{\rm provider}$ and ${\rm CS}$ by ${\hat \theta}_{ij}$, ${\hat R}_{jk}$, ${\hat R}_{ik}$, and ${\hat v}_k$ in similar ways as in (\ref{equ:opprice:payoff}) and (\ref{equ:opprice:cs}). Furthermore, Theorem \ref{theorem:payoff2CS} still holds in the general case, i.e., $\Pi^{\rm provider}=2{\rm CS}$ and both $\Pi^{\rm provider}$ and ${\rm CS}$ are non-decreasing in $a_{xy}$. This is because the properties of effective resistances and ${\bm v}$ used in the proof of Theorem \ref{theorem:payoff2CS} (e.g., local sum rules in (\ref{equ:localsum})) also hold for the modified electrical network and $\left\{{\hat \theta_{ij}}\right\}_{i,j\in{\mathcal N}}$.

\section{Guidelines for Collaboration with Advertisers}\label{sec:guide}
In this section, we provide guidelines for the provider regarding the advertisers to collaborate with. Specifically, we let ${\bm a}\triangleq \left(a_{ij},\left(i,j\right)\in{\mathcal A}\right)$, and use $\Pi^{\rm provider} \left( {\bm a}\right)$ to represent the provider's optimal payoff under the unit ad revenue vector ${\bm a}$. When the provider collaborates with different advertisers and displays different ads in the network, the corresponding values of ${\bm a}$ and $\Pi^{\rm provider} \left( {\bm a}\right)$ are different. We will study the provider's optimal selection of advertisers, the corresponding ${\bm a}$ of which leads to the highest $\Pi^{\rm provider} \left( {\bm a}\right)$.

\begin{figure}[t]
  \centering
  \includegraphics[scale=0.39]{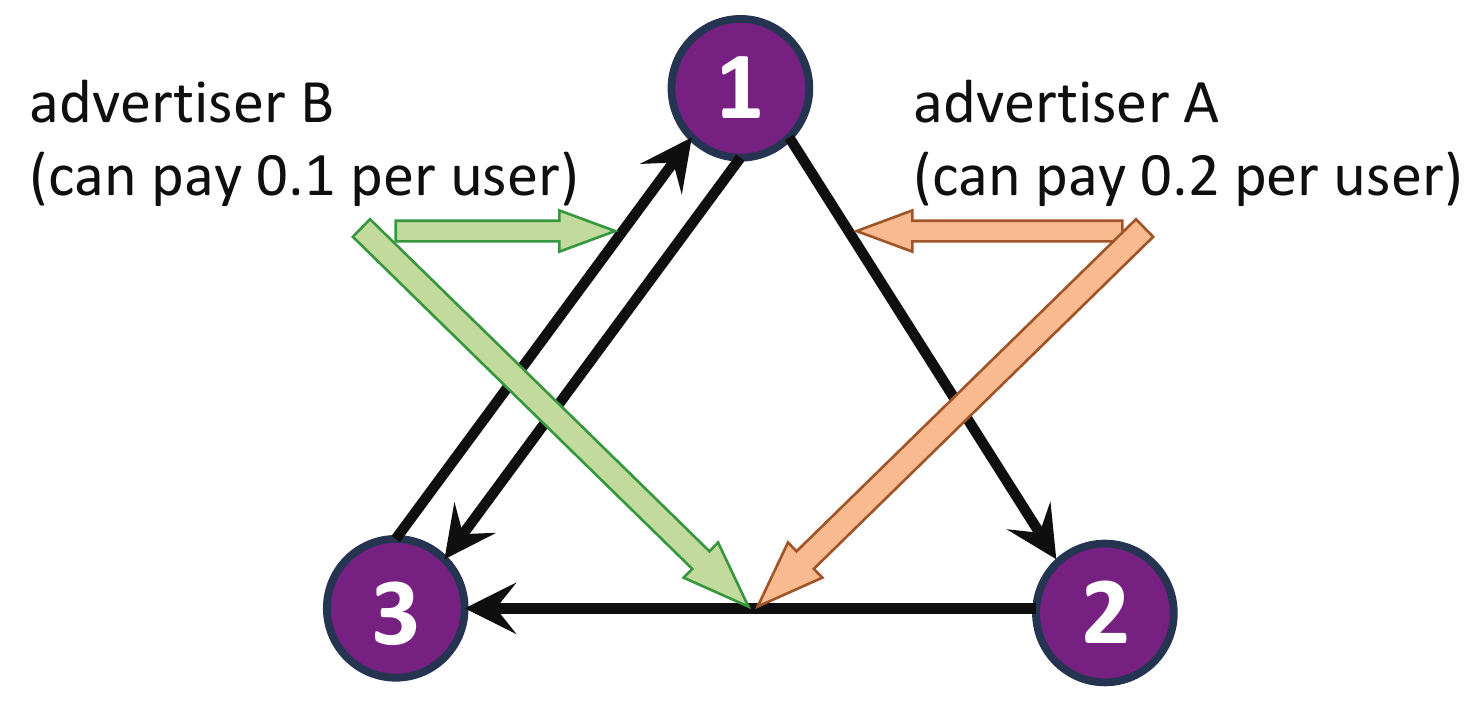}\\
  \caption{An Example of ${\mathcal G}_{\rm d}$.}
  \label{fig:example:a}
\end{figure}

If there is no restriction on the number of advertisers that the provider collaborates with, the provider can decide the ad display for different arcs separately. For an arc, the provider simply displays ads for the advertiser that is willing to pay the most to the provider. {{For example, in Fig. \ref{fig:example:a}, advertiser $A$ is willing to pay $0.2$ if the provider displays its ads to a user on $\left(1,2\right)$ or $\left(2,3\right)$. Advertiser $B$ is willing to pay $0.1$ for advertising to a user on $\left(2,3\right)$ or $\left(3,1\right)$. The provider can collaborate with both advertisers by displaying advertiser $A$'s ads on $\left(1,2\right)$ and $\left(2,3\right)$ and advertiser $B$'s ads on $\left(3,1\right)$. In this case, the corresponding ${\bm a}$ is given by $\left(a_{12},a_{13},a_{23},a_{31}\right)=\left(0.2,0,0.2,0.1\right)$.}}

In practice, it can be time-consuming for the provider to establish collaboration with an advertiser. Hence, the provider may only be able to collaborate with a limited number of advertisers (especially when the provider just starts implementing the in-vehicle advertising). In this case, the provider's ad display decisions for different arcs may be coupled. {{Suppose that the provider can only collaborate with one advertiser in the example in Fig. \ref{fig:example:a}. Collaborating with advertiser $A$ and collaborating with advertiser $B$ correspond to ${\bm a}=\left(0.2,0,0.2,0\right)$ and ${\bm a}=\left(0,0,0.1,0.1\right)$, respectively.}} The provider should solve problem (\ref{opt:sumsub:pricing}) and compute $\Pi^{\rm provider} \left( {\bm a}\right)$ for the two different ${\bm a}$ (if $\mu_{ij}^*=0$ for all $\left(i,j\right)\in{\mathcal A}$, $\Pi^{\rm provider} \left( {\bm a}\right)$ is given by (\ref{equ:opprice:payoff})). Then, the provider can decide its collaborator by comparing $\Pi^{\rm provider} \left( {\bm a}\right)$ under the two ${\bm a}$ vectors. 

For a general network topology and a general parameter setting, when there is a restriction on the number of collaborators, the provider can decide its collaborators based on a similar approach as described above. It can compute $\Pi^{\rm provider} \left( {\bm a}\right)$ for each ${\bm a}$ that it can achieve through collaboration, and then decide the optimal set of collaborators, which leads to the highest $\Pi^{\rm provider} \left( {\bm a}\right)$.

In the rest of this section, we focus on two special cases to study the impact of the network topology on the provider's collaborator selection. In Section \ref{subsec:arcbad}, we focus on the \emph{arc-based advertiser case}, where each advertiser is only willing to advertise on one arc. In Section \ref{subsec:locationbad}, we focus on the \emph{location-based advertiser case}, where each advertiser is only willing to advertise on the arcs pointing to one location. In both cases, we assume that the provider can only collaborate with \emph{one} advertiser, and analyze the provider's optimal choice of this collaborator. 

\subsection{Collaboration with Arc-Based Advertisers}\label{subsec:arcbad}
Suppose that there are $\left|{\mathcal A}\right|$ advertisers and each advertiser wants to advertise on an arc.{\footnote{We consider $\left|{\mathcal A}\right|$ advertisers to simplify the presentation, and we can conduct a similar analysis if the number is not $\left|{\mathcal A}\right|$.}} Furthermore, there do not exist advertisers who want to advertise on the same arc (otherwise, it is trivial to compare these advertisers based on their willingnesses to pay). Hence, each arc $\left(i,j\right)\in{\mathcal A}$ corresponds to an advertiser, and we use $b_{ij}$ to denote its willingness to pay. 

When the provider can only collaborate with one advertiser, its problem is formulated as follows:
\begin{subequations}
\begin{align}
& \max_{\left(x,y\right)\in{\mathcal A}}  {~~}{~~}{~~}\Pi^{\rm provider} \left( {\bm a}\right)\label{equ:arcbad:obj}\\
& {\rm s.t.} {~~}{~~}{~~}{~~}{~~}{~~}{~~}{~~}{~~} a_{ij}=\left\{ {\begin{array}{*{20}{l}}
{b_{ij},}&{{\rm if~}\left(i,j\right)=\left(x,y\right),}\\
{0,}&{{\rm otherwise},}\\
\end{array}} \right. \forall \left(i,j\right)\in{\mathcal A}.\label{equ:arcbad:con}
\end{align}\label{opt:sumsub:arcad}
\end{subequations}
Here, (\ref{equ:arcbad:con}) gives $\bm a$ (i.e., the unit ad revenue vector) when the provider collaborates with the advertiser corresponding to $\left(x,y\right)$. In Section \ref{subsubsec:arcbad:muzero}, we consider the case where $\mu_{ij}^*=0$ for all $\left(i,j\right)\in{\mathcal A}$, and characterize the provider's optimal choice of $\left(x,y\right)$ based on the effective resistances. In Section \ref{subsubsec:arcbad:general}, we consider the general case ($\mu_{ij}^*\ge0$), where we can utilize the results in Section \ref{subsubsec:arcbad:muzero} to reduce the provider's search space of its optimal choice of $\left(x,y\right)$.

\subsubsection{$\mu_{ij}^*=0$ Case}\label{subsubsec:arcbad:muzero}
We consider the case where under each collaboration choice of $\left(x,y\right)$, the corresponding ${\bm a}$ and the other system parameters lead to $\mu_{ij}^*=0$ for all $\left(i,j\right)\in{\mathcal A}$. In this case, $\Pi^{\rm provider} \left( {\bm a}\right)$ is given by (\ref{equ:opprice:payoff}). It is challenging to directly analyze the choice of $\left(x,y\right)$ that maximizes the right side of (\ref{equ:opprice:payoff}). This is because $v_x$ and $v_y$ (defined in (\ref{equ:valueoflocation})) are affected by $a_{xy}$, and the square terms $-\sum_{\left(i,j\right)\in {\mathcal A}} \theta_{ij}\xi_{ij} \left(\frac{\sum_{k\in{\mathcal N}} \left(R_{jk}-R_{ik}\right) v_k}{4 \xi_{ij}}\right)^2$ in (\ref{equ:opprice:payoff}) complicate the coefficients of $v_x$ and $v_y$. The key step of our analysis is to rearrange the right side of (\ref{equ:opprice:payoff}) based on the local sum rules of the resistances in (\ref{equ:localsum}) and get the expression shown in the following proposition. 

\begin{proposition}\label{proposition:deltaa}
The provider's optimal payoff $\Pi^{\rm provider} \left( {\bm a}\right)$ in (\ref{equ:opprice:payoff}) equals $\Delta\left({\bm a}\right)$, which is defined as follows:
\begin{align}
\Delta\left({\bm a}\right) \triangleq  \sum_{\left(i,j\right)\in{\mathcal A}} \theta_{ij} \xi_{ij} \left(\frac{1+a_{ij}-c}{2}\right)^2 -\sum_{\left(i,j\right)\in{\mathcal A}} \left(\frac{1}{8}\theta_{ij} \left(1+a_{ij}-c\right)\sum_{k\in{\mathcal N}} \left(R_{jk}-R_{ik}\right)v_k\right),\label{equ:delta}
\end{align}
where $v_k$ depends on $\bm a$ and is defined in (\ref{equ:valueoflocation}). 
\end{proposition}

For a general parameter setting, we can substitute $\Pi^{\rm provider} \left( {\bm a}\right)=\Delta\left({\bm a}\right)$ into (\ref{equ:arcbad:obj}), and search for the collaboration choice that maximizes $\Delta\left({\bm a}\right)$. In particular, if $\theta_{ij}=\theta_{ji}$ for all $\left(i,j\right)\in{\mathcal A}$, we can simplify the comparison among $\Delta\left({\bm a}\right)$ under different collaboration choices, and have the following result (recall that $\left\{b_{ij}\right\}_{\left(i,j\right)\in{\mathcal A}}$ captures the advertisers' willingnesses to pay).

\begin{theorem}\label{theorem:arcbad}
If $\theta_{ij}=\theta_{ji}$ for all $\left(i,j\right)\in{\mathcal A}$, the provider's optimal collaboration choice is given by
\begin{align}
\left(x,y\right)={\argmax_{\left(i,j\right)\in{\mathcal A}}} {~}\theta_{ij}\xi_{ij} \left(b_{ij}^2+2\left(1-c\right)b_{ij}\right)- \theta_{ij}^2 b_{ij}^2 R_{ij}.\label{equ:theorem:arcbad}
\end{align}
Moreover, $\theta_{ij}\xi_{ij} \left(b_{ij}^2+2\left(1-c\right)b_{ij}\right)- \theta_{ij}^2 b_{ij}^2 R_{ij}$ increases with $b_{ij}$ for all $\left(i,j\right)\in{\mathcal A}$.
\end{theorem}

\begin{figure}[t]
  \centering
  \includegraphics[scale=0.37]{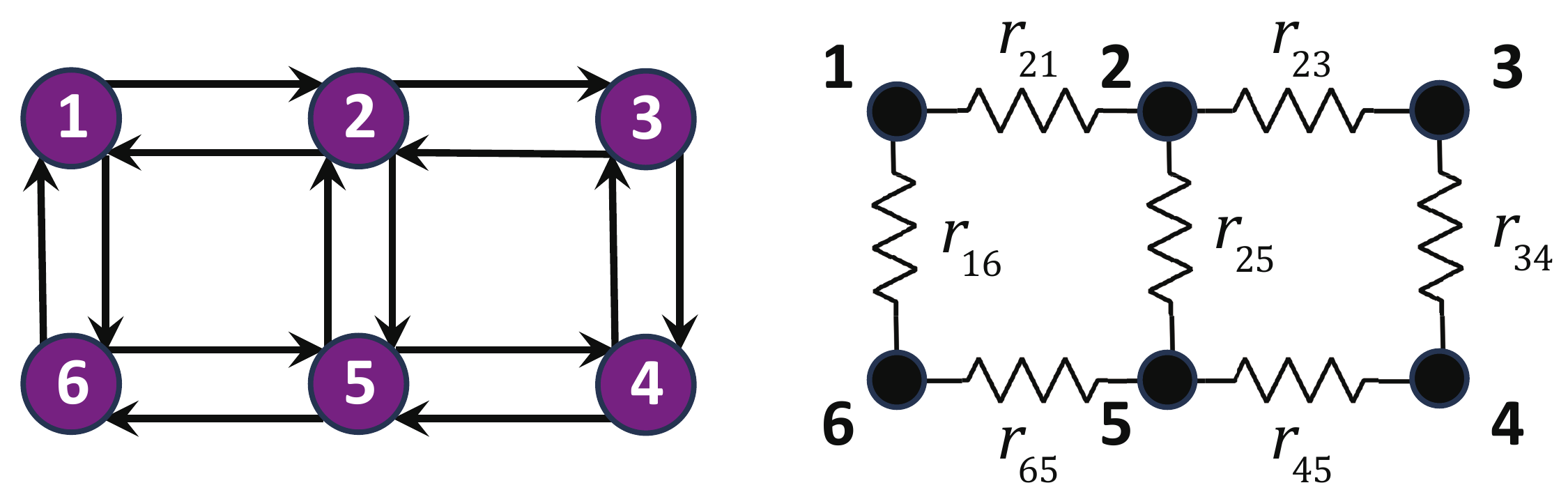}\\
  \caption{An Example of ${\mathcal G}_{\rm d}$ and its electrical network.}
  \label{fig:rectangle}
\end{figure}

{\bf{Remark 4}:} Theorem \ref{theorem:arcbad} implies that when the provider selects one arc to collaborate with the corresponding arc-based advertiser, it needs to consider the following two aspects:
\begin{itemize}
\item First, the mass of the users who want to travel on the selected arc should be large, and the travel time should be long. This enables the provider to get a large total ad revenue by displaying ads on this arc. In (\ref{equ:theorem:arcbad}), this aspect is captured by $\theta_{ij}\xi_{ij} \left(b_{ij}^2+2\left(1-c\right)b_{ij}\right)$, which increases with $\theta_{ij}$ and $\xi_{ij}$ (i.e., the traffic demand and travel time on $\left(i,j\right)$).
\item Second, the provider should be able to route enough vehicles from other arcs to the selected arc. In (\ref{equ:theorem:arcbad}), this aspect is captured by the term $- \theta_{ij}^2 b_{ij}^2 R_{ij}$. Recall that $R_{ij}$ is the effective resistance between $i$ and $j$, and is determined by $\left\{\theta_{ij},\xi_{ij}\right\}_{i,j\in{\mathcal N}}$. Intuitively, a small $R_{ij}$ implies that there are many paths between $i$ and $j$ and these paths have large traffic demand and short travel time.{\footnote{We illustrate an example of ${\mathcal G}_{\rm d}$ and the corresponding electrical network in Fig. \ref{fig:rectangle}. We can see that $R_{25}=\frac{1}{\frac{1}{r_{25}}+\frac{1}{r_{21}+r_{16}+R_{65}}+\frac{1}{r_{23}+r_{34}+r_{45}}}$. Since $r_{21}=\frac{1}{\frac{\theta_{21}}{\xi_{21}}+\frac{\theta_{12}}{\xi_{12}}}$, the value of $R_{25}$ decreases with $\theta_{21}$ and $\theta_{12}$ (i.e., traffic demand), and increases with $\xi_{21}$ and $\xi_{12}$ (i.e., travel time).}} If $\theta_{ij}=\theta_{ji}$ for all $\left(i,j\right)\in{\mathcal A}$, when an arc has a small effective resistance, the provider can route a large mass of vehicles from other arcs to this arc. Hence, the provider prefers to collaborate with the advertiser whose corresponding arc has a small effective resistance.
\end{itemize}
The value of $\theta_{ij}\xi_{ij} \left(b_{ij}^2+2\left(1-c\right)b_{ij}\right)- \theta_{ij}^2 b_{ij}^2 R_{ij}$ increases with $b_{ij}$, which means that an advertiser with a higher willingness to pay is more likely to become the provider's collaborator. 

Next, we consider the situation where the arcs have homogeneous parameters to isolate the impact of the network topology. The following corollary shows that the provider should select the arc with the smallest effective resistance and collaborate with the corresponding advertiser. 

\begin{corollary}\label{corollary:arcbad}
If $\theta_{ij}=\theta_{ji}={\bar \theta}$, $\xi_{ij}={\bar \xi}$, and $b_{ij}={\bar b}$ for all $\left(i,j\right)\in{\mathcal A}$, the provider's optimal collaboration choice is given by $\left(x,y\right)={\argmin_{\left(i,j\right)\in{\mathcal A}}} R_{ij}$.
\end{corollary}
Recall that $R_{ij}=R_{ji}$ for all $i,j\in{\mathcal N}$. When the arcs' parameters are homogeneous, if $\left(x,y\right)$ is an optimal collaboration choice, $\left(y,x\right)$ is also an optimal choice. 

In the example in Fig. \ref{fig:rectangle}, if the traffic demand is ${\bar \theta}$ and the travel time is ${\bar \xi}$ for all arcs in ${\mathcal G}_{\rm d}$, then $R_{25}=R_{52}=\frac{3 {\bar \xi}}{10 {\bar \theta}}$ and the effective resistances of all the other arcs are $\frac{11 {\bar \xi}}{30 {\bar \theta}}$. According to Corollary \ref{corollary:arcbad}, when the arc-based advertisers have the same willingness to pay, the provider should select $\left(2,5\right)$ or $\left(5,2\right)$ to collaborate with the corresponding advertiser. 

\subsubsection{General Case}\label{subsubsec:arcbad:general}
In the general case, under a collaboration choice of $\left(x,y\right)$, the corresponding ${\bm a}$ and the other system parameters may lead to $\mu_{ij}^*>0$ for some $\left(i,j\right)\in{\mathcal A}$. In this case, $\Pi^{\rm provider} \left( {\bm a}\right)$ may not equal $\Delta\left({\bm a}\right)$ in (\ref{equ:delta}). For each collaboration choice of $\left(x,y\right)$, the provider can compute ${\bm a}$ from (\ref{equ:arcbad:con}), and solve problem (\ref{opt:sumsub:pricing}) to get $\Pi^{\rm provider} \left( {\bm a}\right)$. Then, the provider's optimal choice of $\left(x,y\right)$ is the one that achieves the highest $\Pi^{\rm provider} \left( {\bm a}\right)$.

Next, we show that the results in Section \ref{subsubsec:arcbad:muzero} can enable us to reduce the search space of the optimal choice of $\left(x,y\right)$. The provider can perform the following steps:

\emph{Step 1}: For each arc in ${\mathcal A}$, the provider computes ${\bm a}$ from (\ref{equ:arcbad:con}) and further computes the corresponding $\Delta\left({\bm a}\right)$ in (\ref{equ:delta}). Then, the provider sorts all arcs in descending order of their corresponding $\Delta\left({\bm a}\right)$.

\emph{Step 2}: Starting from the first arc (i.e., the one with the largest $\Delta\left({\bm a}\right)$), the provider sequentially checks the arcs. Once the provider finds an arc such that the collaboration with the corresponding advertiser leads to $\mu_{ij}^*=0$ for all $\left(i,j\right)\in{\mathcal A}$, the provider enters Step 3. We assume that this arc has the $h$-th largest $\Delta\left({\bm a}\right)$. If the provider cannot find such an arc after checking all arcs in ${\mathcal A}$, we simply let $h=\left| {\mathcal A} \right|$.

\emph{Step 3}: The provider only needs to compare these $h$ arcs (based on their corresponding $\bm a$ and $\Pi^{\rm provider} \left( {\bm a}\right)$) rather than all arcs in ${\mathcal A}$ to determine the optimal collaboration choice. 

In Appendix \ref{appendix:sec:algorithm}, we prove the optimality of the above steps. The main property used in the proof is that in the general case, we have $\Pi^{\rm provider} \left( {\bm a}\right)\le \Delta\left({\bm a}\right)$.

\subsection{Collaboration with Location-Based Advertisers}\label{subsec:locationbad}
Suppose that there are $N$ advertisers and advertiser $k\in{\mathcal N}$ wants to advertise on the arcs pointing to location $k$. For example, each advertiser plans to advertise a sale at its target location and only display its ads to the users who will arrive at the target location (because of the advertiser's budget constraint). For all $i\in{\mathcal N}$ with $\left(i,k\right)\in{\mathcal A}$, we use $d_{ik}$ to denote advertiser $k$'s willingness to pay if the provider displays its ads to a user on $\left(i,k\right)$. 
We consider $N$ advertisers with different target locations to simplify the presentation, and our analysis can be easily extended to a general situation. 

When the provider can only collaborate with one advertiser, the problem is formulated as follows:
\begin{subequations}
\begin{align}
& \max_{m\in{\mathcal N}}  {~~}{~~}{~~}\Pi^{\rm provider} \left( {\bm a}\right)\label{equ:locationad:obj}\\
& {\rm s.t.} {~~}{~~}{~~}{~~}{~~}{~~} a_{ij}=\left\{ {\begin{array}{*{20}{l}}
{d_{ij},}&{{\rm if~}j=m,}\\
{0,}&{{\rm otherwise},}\\
\end{array}} \right. \forall \left(i,j\right)\in{\mathcal A}.\label{equ:locationad:constraint}
\end{align}\label{opt:sumsub:locationad}
\end{subequations}
Here, (\ref{equ:locationad:constraint}) gives ${\bm a}$ when the provider collaborates with advertiser $m$. Next, we focus on the case where $\mu_{ij}^*=0$ for all $\left(i,j\right)\in{\mathcal A}$. {{The results can be used to reduce the search space of the optimal collaboration choice for the general case ($\mu_{ij}^*\ge0$) in a similar way as in Section \ref{subsubsec:arcbad:general}.}}

When $\mu_{ij}^*=0$ for all $\left(i,j\right)\in{\mathcal A}$, $\Pi^{\rm provider} \left( {\bm a}\right)$ equals $\Delta\left({\bm a}\right)$ in (\ref{equ:delta}). For a general parameter setting, we can solve problem (\ref{opt:sumsub:locationad}) by searching for the collaboration choice that maximizes $\Delta\left({\bm a}\right)$. In particular, if $\theta_{ij}=\theta_{ji}$ for all $\left(i,j\right)\in{\mathcal A}$, we can further simplify the comparison among $\Delta\left({\bm a}\right)$ under different collaboration choices, and have the following result. 

\begin{theorem}\label{theorem:locationad}
If $\theta_{ij}=\theta_{ji}$ for all $\left(i,j\right)\in{\mathcal A}$, the provider's optimal collaboration choice is given by
\begin{align}
m=\argmax_{k\in{\mathcal N}} \!\!\!\! \sum_{s:\left(s,k\right)\in{\mathcal A}} \theta_{sk} \xi_{sk} \left(d_{sk}^2+2\left(1-c\right) d_{sk}\right) \!+\!\! \sum_{s:\left(s,k\right)\in{\mathcal A}} \sum_{t:\left(t,k\right)\in{\mathcal A}} \frac{1}{2} \theta_{sk} \theta_{tk} d_{sk} d_{tk} \left(R_{st}-R_{sk}-R_{tk}\right).\label{equ:theorem:optimalloc}
\end{align}
\end{theorem}

{\bf{Remark 5}:} Theorem \ref{theorem:locationad} characterizes the target location of the advertiser that the provider should collaborate with:
\begin{itemize}
\item First, the location has many incoming arcs, the traffic demand on these arcs is large, and the travel time on these arcs is long. In (\ref{equ:theorem:optimalloc}), these factors are captured by the expression $\sum_{s:\left(s,k\right)\in{\mathcal A}} \theta_{sk} \xi_{sk} \left(d_{sk}^2+2\left(1-c\right) d_{sk}\right)$, which increases with the traffic demand and travel time on each incoming arc of location $k$.
\item Second, the effective resistances of the location's incoming arcs are small, and \emph{the effective resistances between these incoming arcs' origins are large}. In (\ref{equ:theorem:optimalloc}), these are captured by $\sum_{s:\left(s,k\right)\in{\mathcal A}} \sum_{t:\left(t,k\right)\in{\mathcal A}} \frac{1}{2} \theta_{sk} \theta_{tk} d_{sk} d_{tk} \left(-R_{sk}-R_{tk}+R_{st}\right)$, which decreases with $R_{sk}$ and $R_{tk}$ and increases with $R_{st}$. Both $\left(s,k\right)$ and $\left(t,k\right)$ are location $k$'s incoming arcs. As discussed in {Remark 4}, when an arc's effective resistance (e.g., $R_{sk}$ and $R_{tk}$) is small, the provider can route a large mass of vehicles from other arcs to this arc. Furthermore, when $R_{st}$ is large, there are few paths between $s$ and $t$. In this situation, routing more vehicles to $\left(s,k\right)$ does not strongly reduce the mass of vehicles routed to $\left(t,k\right)$, and vice versa. Therefore, when $R_{sk}$ and $R_{tk}$ are small and $R_{st}$ is large, the provider is more likely to collaborate with advertiser $k$.
\end{itemize}

\begin{figure}[t]
  \centering
  \includegraphics[scale=0.37]{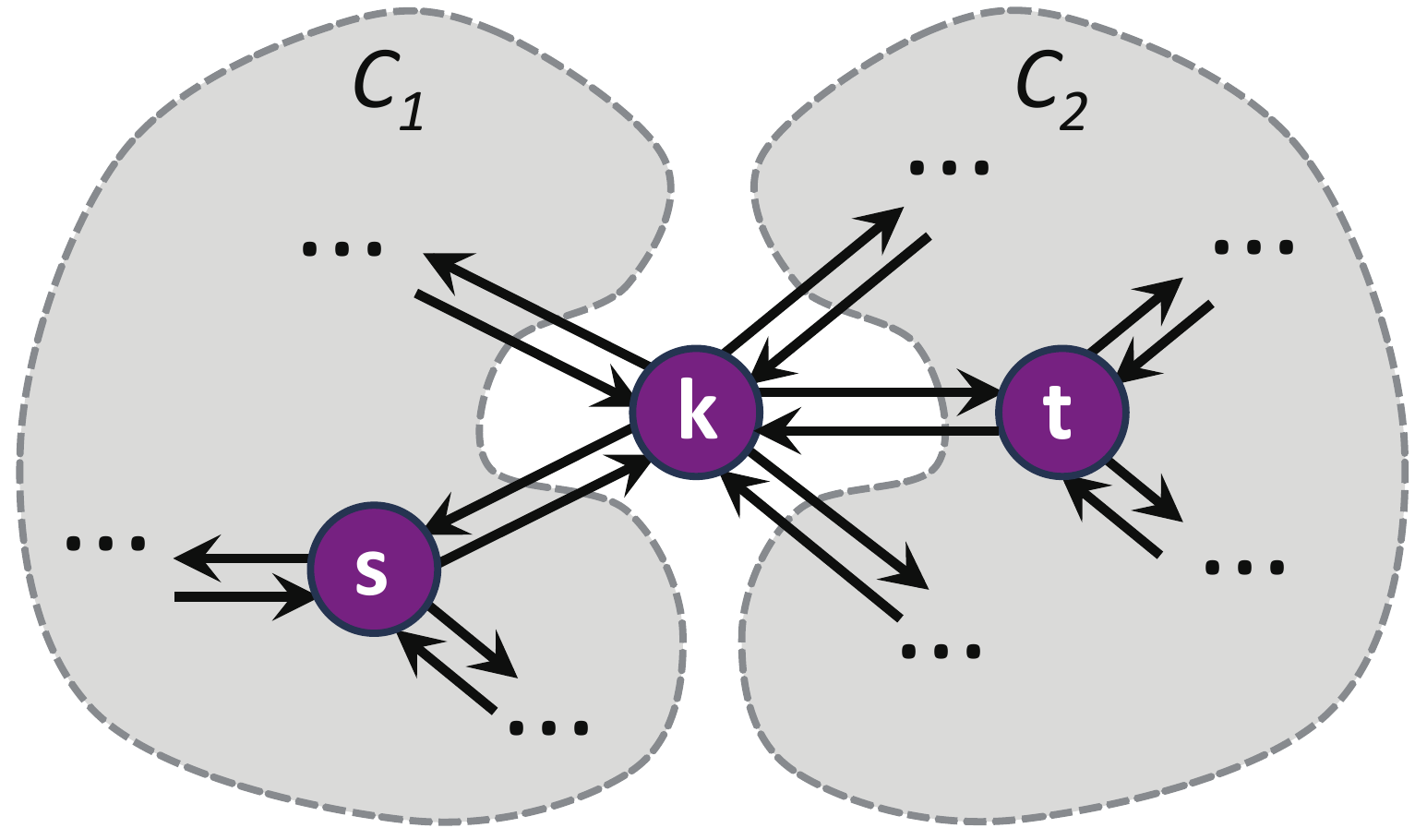}\\
  \caption{An example of ${\mathcal G}_{\rm d}$. If $k$ is a cut-vertex, the effective resistances between its incoming arcs' origins (e.g., $R_{st}$) will be large. In this situation, advertiser $k$ is likely to be the provider's optimal collaborator.}
  \label{fig:locationad}
\end{figure}

We use an example in Fig. \ref{fig:locationad} to illustrate that when $R_{st}$ is large, the provider can route a large mass of vehicles to $\left(s,k\right)$ and $\left(t,k\right)$ at the same time. In Fig. \ref{fig:locationad}, vertex $k$ is a cut-vertex, and $s$ and $t$ belong to different components (i.e., ${\mathcal C}_1$ and ${\mathcal C}_2$) if $k$ is removed. Intuitively, $R_{st}$ is large in this situation. To route vehicles to $\left(s,k\right)$ and $\left(t,k\right)$, the provider should route vehicles from the arcs in ${\mathcal C}_1$ and the arcs in ${\mathcal C}_2$, respectively. Hence, routing more vehicles to $\left(s,k\right)$ and to $\left(t,k\right)$ do not strongly conflict with each other. {{Based on Remark 5 and Fig. \ref{fig:locationad}, the provider can collaborate with a location-based advertiser when its target location has incoming arcs originating from different parts of the network.}}

In Sections \ref{subsec:arcbad} and \ref{subsec:locationbad}, we focus on the situation where the provider can only collaborate with one advertiser to simplify the presentation. {{Suppose that the provider can collaborate with multiple arc-based or location-based advertisers. When $\mu_{ij}^*=0$ for all $\left(i,j\right)\in{\mathcal A}$, the provider's optimal payoff can still be simplified as $\Delta\left({\bm a}\right)$ in (\ref{equ:delta}), and the provider can simply select the collaboration choice that maximizes $\Delta\left({\bm a}\right)$.}}

\section{Model Extensions}\label{sec:numerical}
In Section \ref{subsec:modelextension}, we relax the assumptions made in Section \ref{sec:model}, and introduce an extended model. {{In Section \ref{subsec:numerical:validate}, we validate our results about the provider's advertiser selection choice in the extended model using a real-world dataset.}}

\subsection{Spatial Pricing in An Extended Model}\label{subsec:modelextension}
In Section \ref{sec:model}, we make three main assumptions to simplify the analysis: (i) the users' reservation prices are uniformly distributed; (ii) each vehicle sent on an arc carries a user; (iii) the provider has sufficient vehicles to provide. In the extended model, we relax these assumptions as follows.

First, we use $\theta_{ij} \left(1-F\left(p_{ij}\right)\right)$ to represent the demand function on arc $\left(i,j\right)$, where $F\left(\cdot\right)$ is a non-decreasing function. If $F\left(p_{ij}\right)=\min\left\{p_{ij},1\right\}$, the demand function becomes $\theta_{ij} \max\left\{1-p_{ij},0\right\}$. This corresponds to the setting in Section \ref{sec:model}, where the users have uniformly distributed reservation prices. 

Second, we allow the provider to also route empty vehicles, i.e., the vehicles that do not carry users \cite{braverman2016empty}. Sending empty vehicles from $i$ to $j$ does not directly generate revenue to the provider, but it increases the supply of vehicles at $j$. We use $w_{ij}\ge0$ to denote the mass of empty vehicles that \emph{depart} from $i$ to $j$ in each time slot. The cost of routing an empty vehicle is $\eta c$ (measured in dollars per time slot), where $\eta>0$ denotes the ratio between the costs of routing an empty vehicle and a full vehicle (which carries a user). 

Third, we use $\Psi>0$ to denote the maximum mass of the vehicles (including both full and empty vehicles) that can be sent to the network. 

We formulate the provider's spatial pricing problem in the extended model as follows:
\begin{subequations}
\begin{align}
& \max \sum_{\left(i,j\right)\in{\mathcal A}} \xi_{ij} \theta_{ij} \left(1-F\left(p_{ij}\right)\right)\left(p_{ij}+a_{ij}-c\right)-\sum_{i\in{\mathcal N}} \sum_{j\in{\mathcal N}} \xi_{ij} w_{ij} \eta c \label{optex:obj}\\
& {\rm s.t.}{~~} \sum_{j:\left(i,j\right)\in{\mathcal A}} \theta_{ij} \left(1-F\left(p_{ij}\right)\right) + \sum_{j\in{\mathcal N}} w_{ij} =\sum_{j:\left(j,i\right)\in{\mathcal A}} \theta_{ji} \left(1-F\left(p_{ji}\right)\right) + \sum_{j\in{\mathcal N}} w_{ji}, \forall i\in{\mathcal N},\label{optex:constraint1}\\
& {~~}{~~}{~~}{~~}{~~}{~~}{~~} \sum_{\left(i,j\right)\in{\mathcal A}} \xi_{ij}\theta_{ij} \left(1-F\left(p_{ij}\right)\right)+\sum_{i\in{\mathcal N}} \sum_{j\in{\mathcal N}} \xi_{ij} w_{ij} \le \Psi,\label{optex:constraint2}\\
& {\rm var.}{~~}{~~}{~~}{~~}{~~}p_{ij},\forall \left(i,j\right)\in{\mathcal A}, w_{ij}\ge0, \forall i,j\in{\mathcal N}.\label{optex:var}
\end{align}\label{optex:sumsub}
\end{subequations}
The objective function in (\ref{optex:obj}) captures the provider's time-average payoff. The constraint (\ref{optex:constraint1}) captures the vehicle flow balance at each location $i$ in the presence of empty car routing. The constraint (\ref{optex:constraint2}) is the provider's supply capacity constraint. In any time slot, the mass of full vehicles on arc $\left(i,j\right)$ is $\xi_{ij}\theta_{ij} \left(1-F\left(p_{ij}\right)\right)$,{\footnote{As discussed in Section \ref{subsec:providerpro}, the provider will always choose $p_{ij}$ to ensure that there is no excess demand. As a result, the mass of full vehicles that depart from $i$ to $j$ in each time slot equals the demand (i.e., $\theta_{ij} \left(1-F\left(p_{ij}\right)\right)$). We can multiply it by $\xi_{ij}$ (i.e., travel time) to get the mass of full vehicles on arc $\left(i,j\right)$ in each time slot.}} and the mass of empty vehicles traveling from $i$ to $j$ is $\xi_{ij} w_{ij}$. Constraint (\ref{optex:constraint2}) means that the total mass of vehicles in the network cannot exceed $\Psi$. {{Problem (\ref{optex:sumsub}) can be non-convex, and our analysis of the optimal pricing in Section \ref{sec:prices} cannot be directly applied here.}}

Note that when $F\left(p_{ij}\right)=\min\left\{p_{ij},1\right\}$ and both $\eta$ and $\Psi$ approach infinity, problem (\ref{optex:sumsub}) reduces to problem (\ref{opt:sumsub:pricing}). Specifically, the demand function becomes $\theta_{ij} \max\left\{1-p_{ij},0\right\}$. We can further add the constraint $p_{ij}\le1$, and write the demand as $\theta_{ij} \left(1-p_{ij}\right)$.

\subsection{Numerical Results}\label{subsec:numerical:validate}
We validate our results about the provider's optimal collaboration choice in the extended model. We focus on the provider's collaboration with location-based advertisers. Suppose that the provider can only collaborate with one location-based advertiser. Similar to problem (\ref{opt:sumsub:locationad}), the provider's optimal collaborator selection problem in the extended model is as follows:
\begin{subequations}
\begin{align}
& \max_{m\in{\mathcal N}}  {~~}{~~}{~~}{\tilde \Pi}^{\rm provider} \left( {\bm a}\right)\\
& {\rm s.t.} {~~}{~~}{~~}{~~}{~~}{~~} a_{ij}=\left\{ {\begin{array}{*{20}{l}}
{d_{ij},}&{{\rm if~}j=m,}\\
{0,}&{{\rm otherwise},}\\
\end{array}} \right. \forall \left(i,j\right)\in{\mathcal A},
\end{align}\label{opt:sumsub:locationad:extend}
\end{subequations}
where ${\tilde \Pi}^{\rm provider} \left( {\bm a}\right)$ represents the provider's optimal payoff under the unit ad revenue vector ${\bm a}$ in the extended model. Given ${\bm a}$, ${\tilde \Pi}^{\rm provider} \left( {\bm a}\right)$ equals the optimal objective value of problem (\ref{optex:sumsub}).

According to Section \ref{subsec:locationbad}, in the basic model, the provider can simply select the collaboration choice $m$ that leads to the maximum $\Delta\left({\bm a}\right)$ in (\ref{equ:delta}).{\footnote{Based on our analysis, this strategy is optimal to the provider in the basic model if $\mu_{ij}^*=0$ for all $\left(i,j\right)\in{\mathcal A}$. {{Since this condition usually holds in the numerical experiments, we consider this strategy in this section.}} We will show that this strategy maximizes the provider's payoff for most of the time.}} The numerical results will show that this strategy can also yield a good solution to problem (\ref{opt:sumsub:locationad:extend}). Specifically, we compare the following three advertiser selection strategies: 
\begin{itemize}
\item \emph{Resistance-based selection:} The provider selects the $m$ whose corresponding ${\bm a}$ leads to the maximum $\Delta\left({\bm a}\right)$;
\item \emph{Optimal selection:} The provider selects $m^*$, which is the optimal solution to problem (\ref{opt:sumsub:locationad:extend});
\item \emph{Randomized selection:} The provider randomly selects $m$ from set $\mathcal N$.
\end{itemize}

\begin{figure}[t]
  \centering
  \includegraphics[scale=0.47]{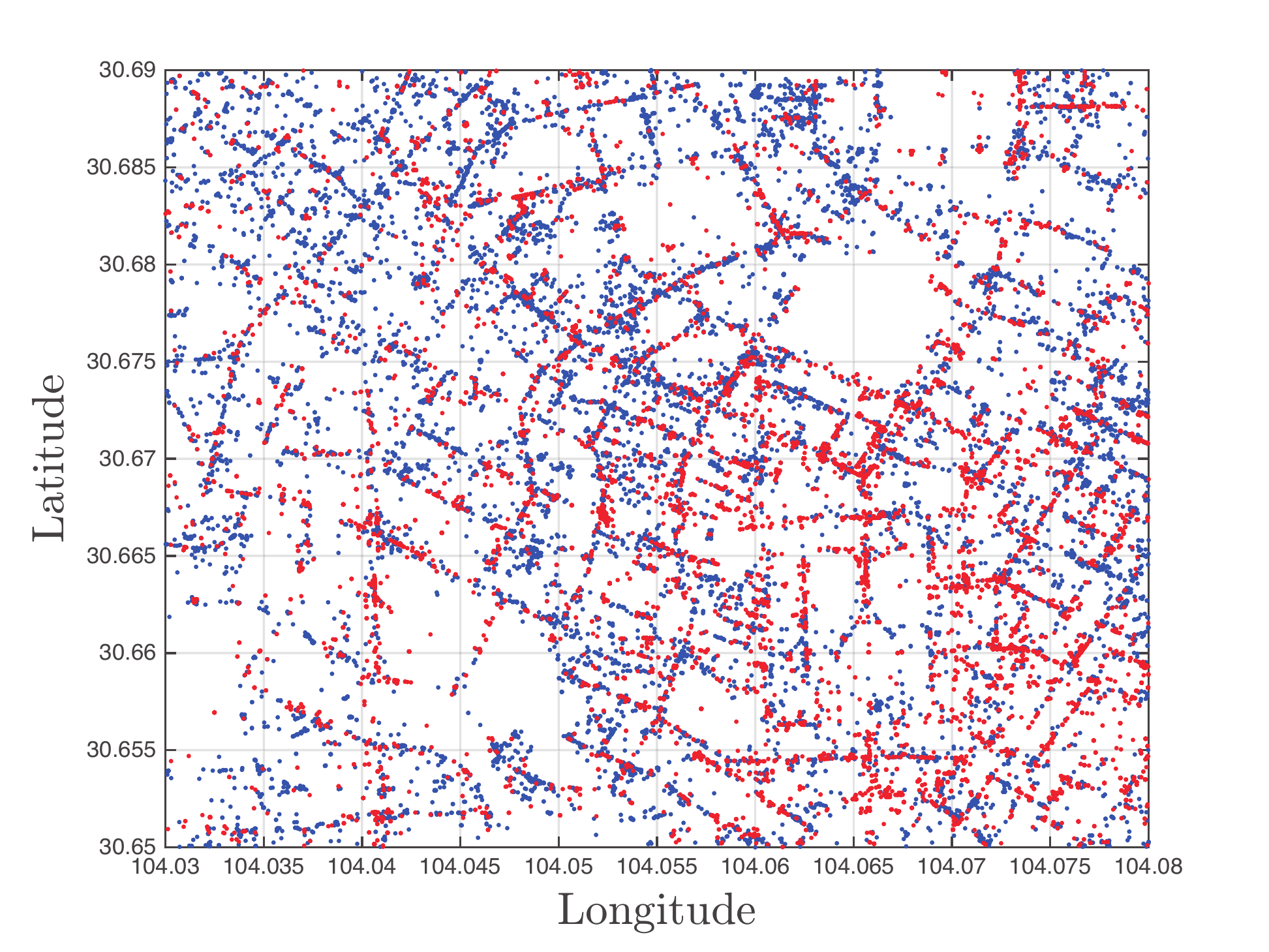}\\
  \caption{Rides' Origins (Blue) and Destinations (Red).}
  \label{appendix:fig:mobi:1}
\end{figure}

We compare the values of ${\tilde \Pi}^{\rm provider} \left( {\bm a}\right)$ achieved by the three advertiser selection strategies. In Sections \ref{subsubsec:extension:uniform} and \ref{subsubsec:extension:exp}, we show the results under different user demand functions.

\begin{figure}[t]
  \centering
  \subfigure[${\tilde \Pi}^{\rm provider}$ Versus $\Psi$ (Uniformly Distributed Reservation Prices).]
  {\label{fig:simu:1}
    \includegraphics[scale=0.33]{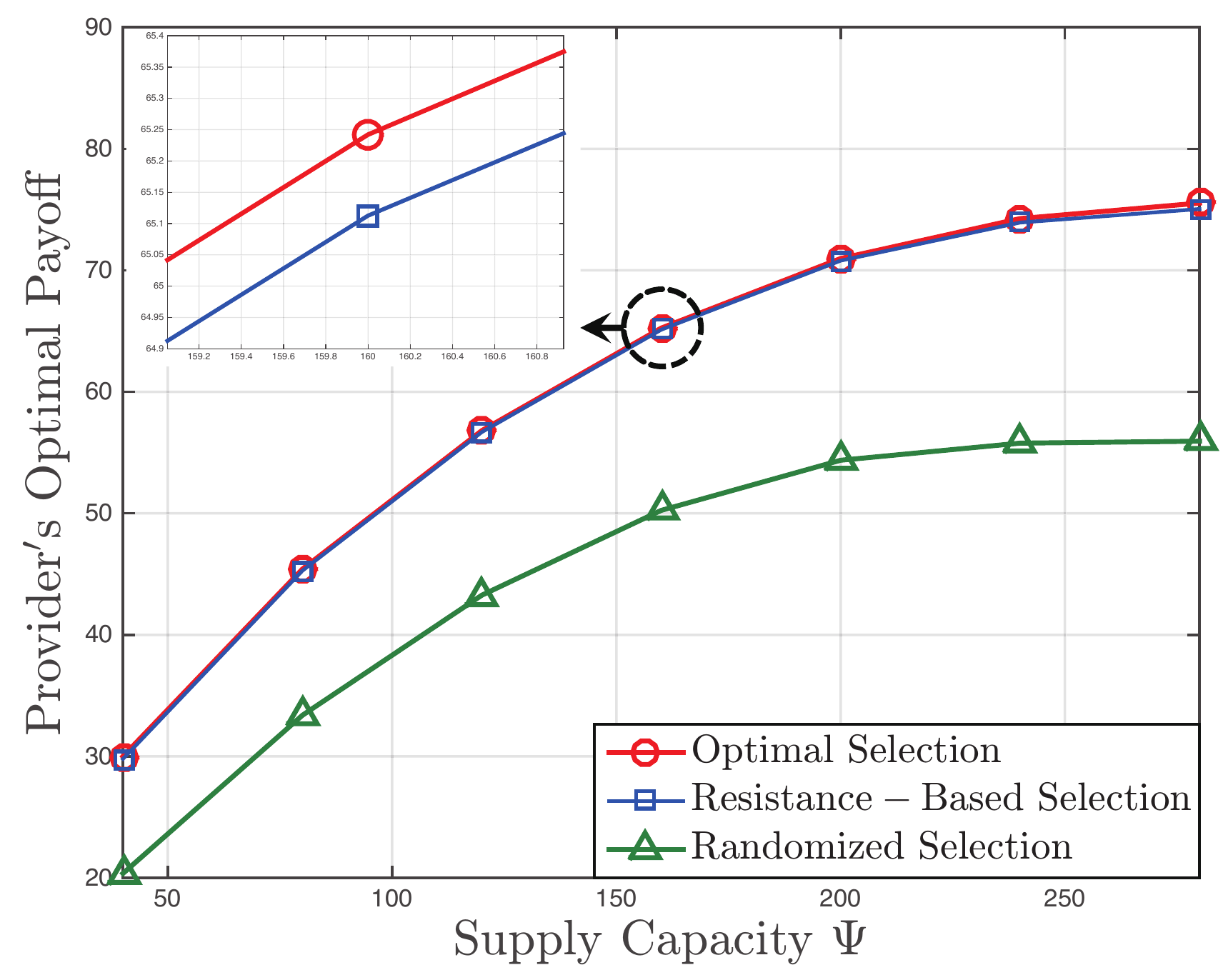}}
  \subfigure[${\tilde \Pi}^{\rm provider}$ Versus $\eta$ (Uniformly Distributed Reservation Prices).]
   {\label{fig:simu:2}
    \includegraphics[scale=0.33]{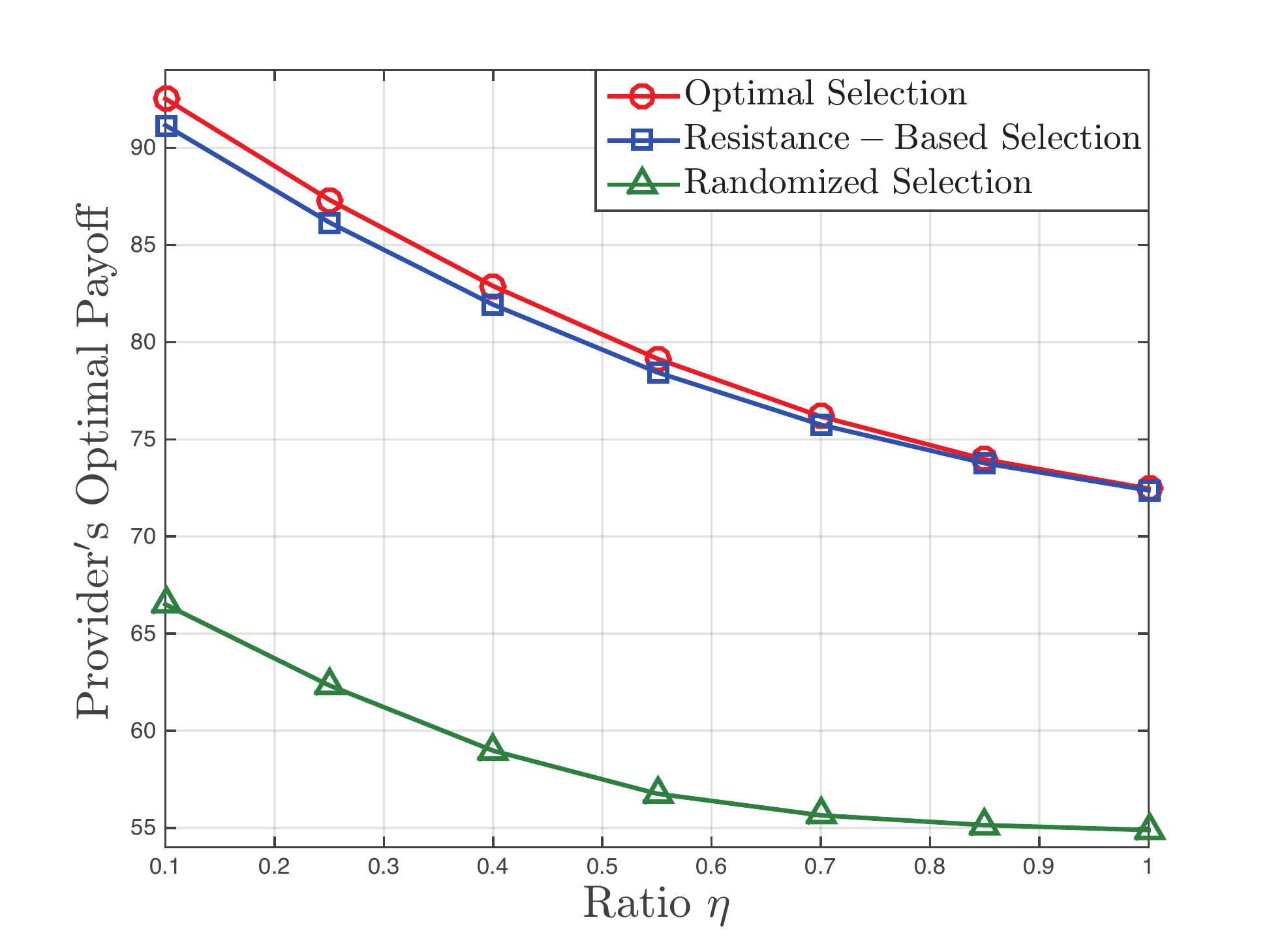}}
  \subfigure[${\tilde \Pi}^{\rm provider}$ Versus $\Psi$ (Exponentially Distributed Reservation Prices).]{
    \label{fig:simu:3}
    \includegraphics[scale=0.33]{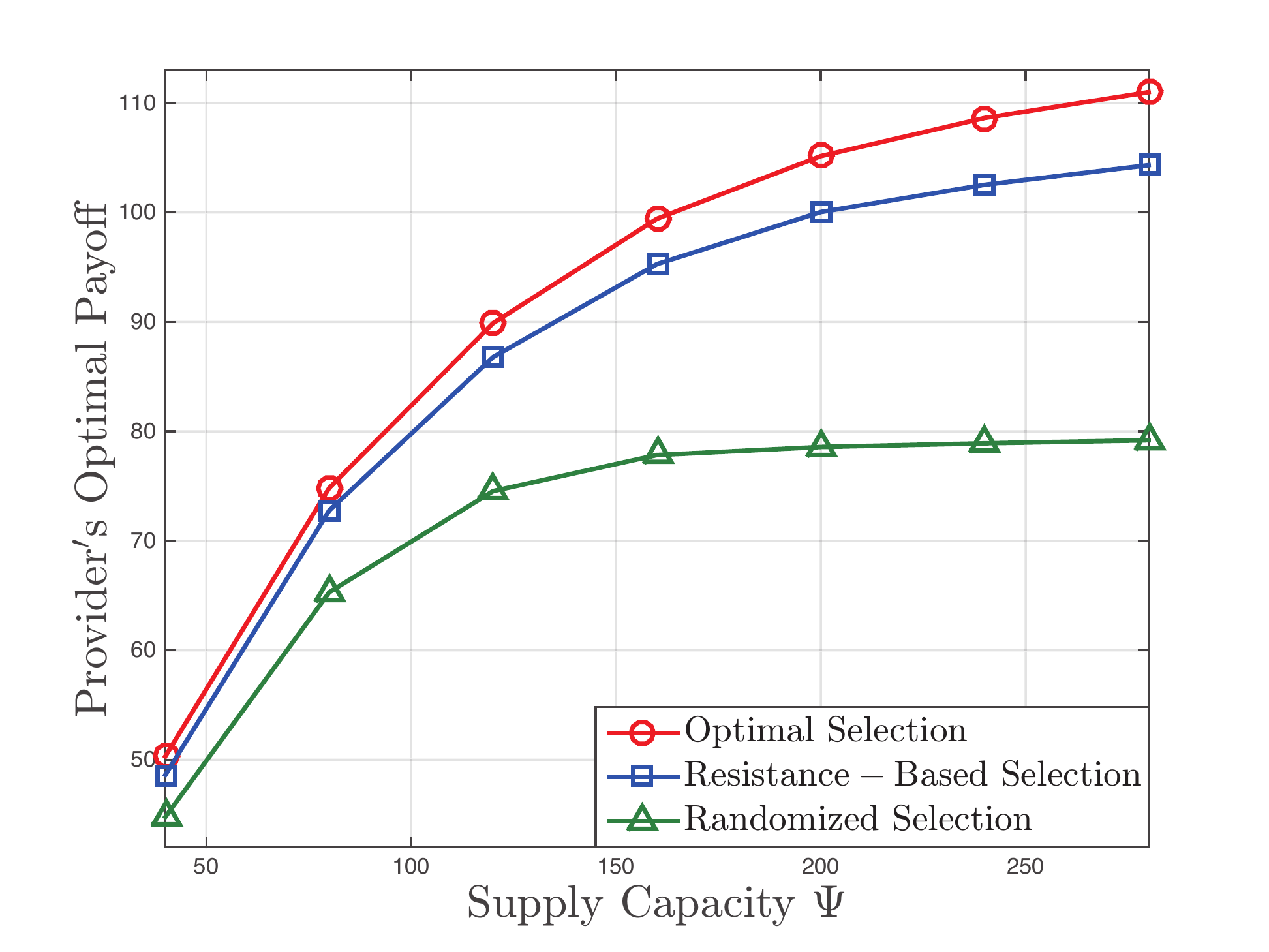}}
  \caption{Comparison Between Advertiser Selection Strategies.}
\end{figure}

\subsubsection{Uniformly Distributed Reservation Prices}\label{subsubsec:extension:uniform}
To focus on the impacts of the supply capacity constraint and the empty car routing, we still consider the uniformly distributed user reservation prices, i.e., we let $F\left(p_{ij}\right)=\min\left\{p_{ij},1\right\}$. 

{{We use a real-world dataset from DiDi Chuxing (the largest ride-sharing platform in China) to generate $\left\{\theta_{ij},\xi_{ij}\right\}_{i,j\in{\mathcal N}}$. The dataset includes information (e.g., origins, destinations, pick-up time, and drop-off time) for the rides taken in November, 2016 in Chengdu, China \cite{DiDidata}. We focus on a $4.8\times4.4 {\rm ~km}^2$ area, which contains both residential and commercial districts. The range of the latitude coordinate of the area is from 30.65 to 30.69, and the range of the longitude coordinate is from 104.03 to 104.08. The commercial districts are mainly in the southeast of the area, and the residential districts are mainly in the northwest. We analyze the rides whose (i) origins and destinations are within this area and (ii) pick-up time and drop-off time are between $7{\rm ~am}$ and $9{\rm ~am}$ on weekdays. In Fig. \ref{appendix:fig:mobi:1}, we plot the origins and destinations of these rides. The origins are in blue and the destinations are in red. Intuitively, most rides are from residential districts to commercial districts during the considered time period. According to the distribution of the rides' origins and destinations, we divide the area into $15$ locations (i.e., $N=15$). For locations $i$ and $j$ ($j\ne i$), we set $\theta_{ij}$ to be the number of rides from $i$ to $j$, and $\xi_{ij}$ to be the average travel time of these rides. We provide the details of the division of the area in Appendix \ref{appendix:sec:mobi:data}.

We set $c=0.6$ and $\eta=0.8$, and randomly choose the value for $d_{ij}$ based on an exponential distribution (with mean equal to $0.4$). We compare the provider's payoffs achieved by the three advertiser selection strategies under different vehicle supply capacity $\Psi$. We change $\Psi$ from $40$ to $280$, and run the experiment $50$ times for each value of $\Psi$. In Fig. \ref{fig:simu:1}, we plot the provider's payoff against $\Psi$. First, we observe that the provider's payoff is always non-decreasing in $\Psi$. This is intuitive, since the provider can earn more revenue when it has more vehicles to provide. Second, the gap between the payoffs under our resistance-based selection (blue curve) and the optimal selection (red curve) is less than $0.7\%$ of the payoff under the optimal selection. Our resistance-based selection leads to the same advertiser collaboration choice as the optimal selection for most of the time, and hence achieves a close-to-optimal provider's payoff. Third, our resistance-based selection achieves a much higher payoff than the randomized selection (green curve). For example, when $\Psi=280$, {{the resistance-based selection improves the provider's payoff over the randomized selection by $34.2\%$}}. 

Then, we compare the three selection strategies under different $\eta$ (i.e., the ratio between the costs of routing an empty vehicle and a full vehicle). We let $\Psi=300$, and the settings of $c$ and $d_{ij}$ are the same as those in Fig. \ref{fig:simu:1}. We change $\eta$ from $0.1$ to $1$, and run the experiment $50$ times for each value of $\eta$. In Fig. \ref{fig:simu:2}, we plot the provider's payoff against $\eta$. First, we can see that the provider's payoff is non-increasing in $\eta$. This is because it is costly to route empty vehicles under a large $\eta$. Second, our resistance-based selection achieves a close-to-optimal provider's payoff, and outperforms the randomized selection, which is similar to the result in Fig. \ref{fig:simu:1}.}}

\subsubsection{Exponentially Distributed Reservation Prices}\label{subsubsec:extension:exp}
Next, we verify the observation in Section \ref{subsubsec:extension:uniform} by considering exponentially distributed user reservation prices.{\footnote{{Besides the uniform distribution and exponential distribution, some other distributions have been used to model the distribution of users' reservation prices. For example, reference \cite{banerjee2015pricing} considered a general Gamma distribution.}}} We let $F\left(p_{ij}\right)=1-e^{-\gamma p_{ij}}$ and choose $\gamma=2$. This implies that the reservation prices of the $\theta_{ij}$ users on $\left(i,j\right)\in{\mathcal A}$ follow an exponential distribution with mean equal to $0.5$. In \cite{fang2017prices}, real data shows that the exponential function can well approximate the actual user demand function for a ride-sharing platform's service. Reference \cite{ma2018spatio} also modeled riders' demand function by an exponential function.

{{We compare the three advertiser selection strategies under different $\Psi$ (the results of their comparison under different $\eta$ are similar to those in Fig. \ref{fig:simu:2}, and hence are omitted). The parameter settings are the same as those in Fig. \ref{fig:simu:1}. In Fig. \ref{fig:simu:3}, we plot the provider's payoff against $\Psi$. We can see that the gap between the payoffs under our resistance-based selection and the optimal selection is greater than that in the uniformly distributed reservation price case (shown in Fig. \ref{fig:simu:1}). The gap is up to $6\%$ of the payoff under the optimal selection. 
However, our resistance-based selection has a much lower computation complexity. It only requires the provider to compute $\Delta\left({\bm a}\right)$ (which has a closed-form expression in (\ref{equ:delta})) for each collaboration choice and do the comparison. In contrast, the optimal selection requires the provider to compute ${\tilde \Pi}^{\rm provider} \left( {\bm a}\right)$ (i.e., the optimal objective value of problem (\ref{optex:sumsub})) for each collaboration choice and do the comparison. In each of our experiments, the time required for our resistance-based selection and the optimal selection to determine the advertiser selection choice in MATLAB are around 0.2 seconds and 9 minutes, respectively. Fig. \ref{fig:simu:3} also shows that the resistance-based selection outperforms the randomized selection. When $\Psi=280$, the resistance-based selection improves the provider's payoff over the randomized selection by $31.7\%$. 

From Fig. \ref{fig:simu:1}, \ref{fig:simu:2}, and \ref{fig:simu:3}, our resistance-based advertiser selection achieves a high provider's payoff in the extended model. Under the resistance-based selection, the provider selects the collaborator whose corresponding ${\bm a}$ leads to the maximum $\Delta\left({\bm a}\right)$. Therefore, the numerical results imply that the comparison among ${\tilde \Pi}^{\rm provider} \left( {\bm a}\right)$ under different collaboration choices can be well approximated by the comparison among $\Delta\left({\bm a}\right)$. Recall that in the basic model, we have used $\Delta\left({\bm a}\right)$ to derive many insights about the provider's collaboration with the advertisers (e.g., Remarks 4 and 5). Since the comparison of $\Delta\left({\bm a}\right)$ can well approximate the comparison of ${\tilde \Pi}^{\rm provider} \left( {\bm a}\right)$, these insights also hold in the extended model.}}

\section{Conclusion}
In-vehicle advertising is a promising approach to monetize a vehicle service, especially an autonomous vehicle service. We analyzed in-vehicle advertising's economic impact based on effective resistances, and derived the following insights. 
First, if an arc has a long travel time, the network topology and other arcs' ad revenues have a small impact on the arc's optimal price. 
Second, increasing an arc's ad revenue may reduce some arcs' consumer surplus, but does not reduce the total consumer surplus. 
Third, the provider should collaborate with a location-based advertiser whose target location has incoming arcs originating from locations in different parts of the network. 

{{There are some limitations to our study, which may open up directions for future research. First, we considered time-invariant user demand and travel time. One extension is to model the system by a closed-queueing network \cite{braverman2016empty,banerjee2016pricing}, where the arrival of user requests on an arc follows a Poisson process and the travel time follows an exponential distribution. Given a pricing and vehicle assignment policy, the number of vehicles on different arcs can be characterized by a continuous time Markov chain. As studied in \cite{braverman2016empty}, we may prove that the closed-queueing network can be approximated by a fluid model under a fixed pricing and vehicle assignment policy. Therefore, we can leverage the fluid model to design the pricing and vehicle assignment policy in the presence of ad revenues. Second, our work focuses on a monopolistic provider. When multiple providers offer the vehicle service, users choose the providers based on the providers' service prices and vehicle availabilities. Each provider decides its pricing and vehicle assignment by anticipating users' choices. Since each provider needs to consider its own vehicle flow balance constraints at all locations, the analysis of providers' decisions in the equilibrium is challenging. If the providers have limited vehicle supplies, they may offer the service in different regions in the equilibrium to reduce their competition. It is also interesting to study the impact of the in-vehicle advertising on the providers' competition. Third, we did not consider the influence of vehicle assignment on the travel time on different arcs. In practice, when the number of vehicles on an arc is large, the corresponding roads can be congested, which may lead to a large travel time on the arc. Fourth, we investigated a general vehicle service provider in this work. We can focus on a ride-sharing platform and analyze its spatial compensation to drivers considering the ad revenues.}}

\begin{acks}
This research was supported in part by NSF grants TWC-1314620, AST-1547328 and CNS-1701921.
\end{acks}

\bibliographystyle{ACM-Reference-Format}
\bibliography{adflow}

\appendix

\section{Proof of Proposition \ref{proposition:KKTprice}}
\begin{proof}
The objective function in (\ref{opt2:objective}) is quadratic and concave, and the constraints (\ref{opt2:constraint:flow}) are affine. In this case, problem (\ref{opt:sumsub:pricing}) is convex, and the KKT conditions are necessary and sufficient for optimality. Based on the stationarity condition, we have the following relation:
\begin{align}
\theta_{ij}\xi_{ij}  \left(2p_{ij}^*-1-c+a_{ij} \right)-\lambda_i^* \theta_{ij} +\lambda_j^* \theta_{ij}+\mu_{ij}^*=0,\forall \left(i,j\right)\in{\mathcal A}.
\end{align}
After rearrangement, we can get the expression of $p_{ij}^*$:
\begin{align}
p_{ij}^*=\frac{1-a_{ij}+c}{2}+\frac{\lambda_i^*-\lambda_j^*}{2\xi_{ij}}-\frac{\mu_{ij}^*}{2\xi_{ij}\theta_{ij}},\forall \left(i,j\right)\in{\mathcal A}.
\end{align}
\end{proof}

\section{Proof of Proposition \ref{proposition:solutionspace}}
\begin{proof}
First, we show that ${\bm \lambda}={\bm L}^+ {\bm v}$ is one solution to ${\bm L} {\bm \lambda}={\bm v}$. Given the Laplacian matrix ${\bm L}$ and its generalized inverse ${\bm L}^+$, we have the following property \cite{bozzo2012effective}:
\begin{align}
{\bm L} {\bm L}^+ = {\bm I} - \frac{1}{N} {\bm J},
\end{align}
where ${\bm I}$ is the $N\times N$ identity matrix and ${\bm J}$ is the $N\times N$ all-ones matrix (whose entries are $1$). Then, we can see that 
\begin{align}
{\bm L} {\bm L}^+ {\bm v}= {\bm I}{\bm v} - \frac{1}{N} {\bm J}{\bm v} ={\bm v},\label{appendix:equ:lapl}
\end{align}
where we use $\sum_{k\in{\mathcal N}} v_k=0$ (which is obtained based on ${\bm v}$'s definition in (\ref{equ:valueoflocation})) to derive the second equality. According to (\ref{appendix:equ:lapl}), ${\bm \lambda}={\bm L}^+ {\bm v}$ is one solution to ${\bm L} {\bm \lambda}={\bm v}$.

Second, we prove that the solution space of ${\bm L} {\bm \lambda}={\bm v}$ is $\left\{{\bm \lambda}: {\bm \lambda}={\bm L}^+ {\bm v} + \beta \left(1,1,\ldots,1\right)^\intercal, \beta\in{\mathbb R}\right\}$. Let ${\bm \lambda}_1$ and ${\bm \lambda}_2$ be two solutions to ${\bm L} {\bm \lambda}={\bm v}$. We can see that ${\bm L} \left({\bm \lambda}_1-{\bm \lambda}_2\right)={\bm 0}$. Since ${\bm L}$ is the Laplacian matrix, its null space is one-dimensional and it satisfies ${\bm L} \left(1,1,\ldots,1\right)^\intercal={\bm 0}$. Therefore, vector $\left(1,1,\ldots,1\right)^\intercal$ belongs to the null space, and there exists $\beta\in{\mathbb R}$ such that ${\bm \lambda}_1-{\bm \lambda}_2=\beta \left(1,1,\ldots,1\right)^\intercal$. Because the result holds for any two solutions of ${\bm L} {\bm \lambda}={\bm v}$, the solution space of ${\bm L} {\bm \lambda}={\bm v}$ is characterized by $\left\{{\bm \lambda}: {\bm \lambda}={\bm L}^+ {\bm v} + \beta \left(1,1,\ldots,1\right)^\intercal, \beta\in{\mathbb R}\right\}$.

Since any solution of ${\bm L} {\bm \lambda}={\bm v}$ can be written in the form of ${\bm \lambda}={\bm L}^+ {\bm v} + \beta \left(1,1,\ldots,1\right)^\intercal$ for some $\beta\in{\mathbb R}$, the value of $\lambda_i-\lambda_j$ ($i,j\in{\mathcal N}$) equals $\sum_{k\in{\mathcal N}} \left(l_{ik}^+ - l_{jk}^+\right) v_k$.
\end{proof}

\section{Proof of Theorem \ref{theorem:priceresistance}}
\begin{proof}
From (\ref{equ:RandLplus}), we have the following relations for all $i,j,k\in{\mathcal N}$:
\begin{align}
& l_{ik}^+ =\frac{l_{ii}^+ + l_{kk}^+ - R_{ik}}{2},\\
& l_{jk}^+ =\frac{l_{jj}^+ + l_{kk}^+ - R_{jk}}{2}.
\end{align}
According to (\ref{equ:gaplambda}), if $\mu_{ij}^*=0$ for all $\left(i,j\right)\in{\mathcal A}$, then $\lambda_i^*-\lambda_j^*=\sum_{k\in{\mathcal N}} \left(l_{ik}^+ - l_{jk}^+\right) v_k$ for all $i,j\in{\mathcal N}$. By substituting the expressions of $l_{ik}^+$ and $l_{jk}^+$ into the expression of $\lambda_i^*-\lambda_j^*$, we get the following result for all $i,j\in{\mathcal N}$:
\begin{align}
\lambda_i^*-\lambda_j^*=\sum_{k\in{\mathcal N}} \left(\frac{l_{ii}^+ + l_{kk}^+ - R_{ik}}{2}  -\left(\frac{l_{jj}^+ + l_{kk}^+ - R_{jk}}{2}\right) \right) v_k = \sum_{k\in{\mathcal N}} \left(\frac{l_{ii}^+  - R_{ik} -  l_{jj}^+  + R_{jk}}{2}  \right) v_k.
\end{align}
Because $\sum_{k\in{\mathcal N}} v_k=0$, we have $\sum_{k\in{\mathcal N}} l_{ii}^+ v_k=0$ and $\sum_{k\in{\mathcal N}} l_{jj}^+ v_k=0$. Hence, the value of $\lambda_i^*-\lambda_j^*$ ($i,j\in{\mathcal N}$) is given by
\begin{align}
\lambda_i^*-\lambda_j^* =\frac{1}{2}\sum_{k\in{\mathcal N}} \left(R_{jk}-R_{ik}\right)v_k.
\end{align}
When $\mu_{ij}^*=0$ for all $\left(i,j\right)\in{\mathcal A}$, we substitute the above expression into (\ref{equ:optimalp}), and have
\begin{align}
p_{ij}^*=\frac{1-a_{ij}+c}{2}+\frac{1}{4\xi_{ij}} \sum_{k\in{\mathcal N}} \left(R_{jk}-R_{ik}\right) v_k,\forall \left(i,j\right)\in{\mathcal A}.
\end{align}
\end{proof}

\section{Proof of Proposition \ref{proposition:asymbound}}
\begin{proof}
We first prove that when the system parameters satisfy the condition $\sum_{k\in{\mathcal N}} \left|v_k\right| \le \min_{\left(x,y\right)\in{\mathcal A}} 2\left(\theta_{xy}+\theta_{yx}\frac{\xi_{xy}}{\xi_{yx}}\right)\left(1+a_{xy}-c\right)$, the following inequality holds for all $\left(i,j\right)\in{\mathcal A}$:
\begin{align}
\frac{1-a_{ij}+c}{2}+\frac{1}{4\xi_{ij}} \sum_{k\in{\mathcal N}} \left(R_{jk}-R_{ik}\right) v_k\le1.
\end{align}
Because the effective resistances satisfy the triangle inequality \cite{dorfler2018electrical}, we have $R_{jk}-R_{ik}\le R_{ij}$ and $R_{ik}-R_{jk}\le R_{ij}$. Moreover, the effective resistance between two locations is always no greater than the resistance of the resistor between them, which implies that $R_{ij}\le r_{ij}$. Since $r_{ij}=\frac{1}{\frac{\theta_{ij}}{\xi_{ij}}+\frac{\theta_{ji}}{\xi_{ji}}}$, the following inequality holds:
\begin{align}
\left| R_{jk}-R_{ik} \right| \le \frac{1}{\frac{\theta_{ij}}{\xi_{ij}}+\frac{\theta_{ji}}{\xi_{ji}}}.
\end{align}
Using this inequality, we can further see that
\begin{align}
\nonumber
& \frac{1-a_{ij}+c}{2}+\frac{1}{4\xi_{ij}} \sum_{k\in{\mathcal N}} \left(R_{jk}-R_{ik}\right) v_k \\
\nonumber
& \le \frac{1-a_{ij}+c}{2}+\frac{1}{4\xi_{ij}} \sum_{k\in{\mathcal N}} \left| R_{jk}-R_{ik}\right| \left| v_k \right| \\
& \le \frac{1-a_{ij}+c}{2}+\frac{1}{4\xi_{ij}} \frac{1}{\frac{\theta_{ij}}{\xi_{ij}}+\frac{\theta_{ji}}{\xi_{ji}}} \sum_{k\in{\mathcal N}} \left| v_k \right|. \label{appendix:equ:mubound}
\end{align}
When $\sum_{k\in{\mathcal N}} \left|v_k\right| \le \min_{\left(x,y\right)\in{\mathcal A}} 2\left(\theta_{xy}+\theta_{yx}\frac{\xi_{xy}}{\xi_{yx}}\right)\left(1+a_{xy}-c\right)$, the right side of (\ref{appendix:equ:mubound}) is upper bounded as follows:
\begin{align}
\nonumber
& \frac{1-a_{ij}+c}{2}+\frac{1}{4\xi_{ij}} \frac{1}{\frac{\theta_{ij}}{\xi_{ij}}+\frac{\theta_{ji}}{\xi_{ji}}} \sum_{k\in{\mathcal N}} \left| v_k \right| \\
\nonumber
& \le \frac{1-a_{ij}+c}{2}+\frac{1}{2\xi_{ij}} \frac{1}{\frac{\theta_{ij}}{\xi_{ij}}+\frac{\theta_{ji}}{\xi_{ji}}} \min_{\left(x,y\right)\in{\mathcal A}} \left(\theta_{xy}+\theta_{yx}\frac{\xi_{xy}}{\xi_{yx}}\right)\left(1+a_{xy}-c\right)\\
\nonumber
& \le \frac{1-a_{ij}+c}{2}+\frac{1}{2} \frac{1}{ \theta_{ij}+\theta_{ji}\frac{\xi_{ij} }{\xi_{ji}}}  \left(\theta_{ij}+\theta_{ji}\frac{\xi_{ij}}{\xi_{ji}}\right)\left(1+a_{ij}-c\right)\\
& = 1.\label{appendix:equ:mubound2}
\end{align}
Combining (\ref{appendix:equ:mubound}) and (\ref{appendix:equ:mubound2}), we can prove that if $\sum_{k\in{\mathcal N}} \left|v_k\right| \!\le\! \min_{\left(x,y\right)\in{\mathcal A}} 2\left(\theta_{xy}+\theta_{yx}\frac{\xi_{xy}}{\xi_{yx}}\right)\left(1+a_{xy}-c\right)$, the following inequality holds for all $\left(i,j\right)\in{\mathcal A}$:
\begin{align}
\frac{1-a_{ij}+c}{2}+\frac{1}{4\xi_{ij}} \sum_{k\in{\mathcal N}} \left(R_{jk}-R_{ik}\right) v_k\le1.\label{appendix:equ:inequality}
\end{align}

Then, we let $p_{ij}= \frac{1-a_{ij}+c}{2}+\frac{1}{4\xi_{ij}} \sum_{k\in{\mathcal N}} \left(R_{jk}-R_{ik}\right) v_k$ and $\mu_{ij}=0$ for all $\left(i,j\right)\in{\mathcal A}$, and let ${\bm \lambda}$ be the  solution to ${\bm L} {\bm \lambda}={\bm v}$. From (\ref{appendix:equ:inequality}), $p_{ij}\le 1$ for all $\left(i,j\right)\in{\mathcal A}$. We can verify that $\left\{p_{ij},\mu_{ij}\right\}_{\left(i,j\right)\in{\mathcal A}}$ and ${\bm \lambda}$ satisfy the KKT conditions. Hence, they constitute an optimal solution to problem (\ref{opt:sumsub:pricing}).
\end{proof}

\section{Proof of Theorem \ref{theorem:adarc}}
\begin{proof}
From (\ref{equ:priceresistance}), the optimal price for arc $\left(i,j\right)$ is as follows:
\begin{align}
p_{ij}^*=\frac{1-a_{ij}+c}{2}+\frac{1}{4\xi_{ij}} \sum_{k\in{\mathcal N}} \left(R_{jk}-R_{ik}\right) v_k,\forall \left(i,j\right)\in{\mathcal A},
\end{align}
where $v_k$ is given by
\begin{align}
v_k = \sum_{m:\left(k,m\right)\in{\mathcal A}} \theta_{km} \left(1+a_{km}-c\right)- \sum_{m:\left(m,k\right)\in{\mathcal A}} \theta_{mk} \left(1+a_{mk}-c\right).
\end{align}
By taking the derivative of $p_{ij}^*$ with respect to $a_{xy}$, we can get the following relation:
\begin{align}
\frac{\partial p_{ij}^*}{\partial a_{xy}}=-\frac{1}{2}{\mathbbm 1}_{\left\{i=x,j=y\right\}}+\frac{\theta_{xy}}{4\xi_{ij}}\left(R_{jx}-R_{ix}-R_{jy}+R_{iy}\right).
\end{align}
In particular, when $i=x$ and $j=y$, the above equality becomes
\begin{align}
\frac{\partial p_{xy}^*}{\partial a_{xy}} =-\frac{1}{2}+\frac{\theta_{xy}}{4\xi_{xy}}\left(R_{yx}-R_{xx}-R_{yy}+R_{xy}\right)= -\frac{1}{2}+\frac{\theta_{xy}}{2\xi_{xy}} R_{xy},
\end{align}
where the second equality is based on $R_{xx}=R_{yy}=0$ and $R_{xy}=R_{yx}$. The effective resistance between two locations is always no greater than the resistance of the resistor between them. Hence, $R_{xy}\le r_{xy}=\frac{1}{\frac{\theta_{xy}}{\xi_{xy}}+\frac{\theta_{yx}}{\xi_{yx}}}$. Using this inequality, we can see that
\begin{align}
\frac{\partial p_{xy}^*}{\partial a_{xy}}\le -\frac{1}{2}+\frac{\theta_{xy}}{2\xi_{xy}} \frac{1}{\frac{\theta_{xy}}{\xi_{xy}}+\frac{\theta_{yx}}{\xi_{yx}}}\le0.
\end{align}
\end{proof}

\section{Proof of Corollary \ref{corollary:comnode}}
\begin{proof}
When $i=x$ and $j\ne y$, Eq. (\ref{equ:impactadprice}) becomes
\begin{align}
\frac{\partial p_{xj}^*}{\partial a_{xy}}=\frac{\theta_{xy}}{4\xi_{xj}}\left(R_{jx}-R_{xx}-R_{jy}+R_{xy}\right)=\frac{\theta_{xy}}{4\xi_{xj}}\left(R_{jx}-R_{jy}+R_{xy}\right).
\end{align}
The second equality is based on $R_{xx}=0$. According to the triangle inequality of effective resistances \cite{dorfler2018electrical}, we have $R_{jx}-R_{jy}+R_{xy}\ge0$. Hence, the value of $\frac{\partial p_{xj}^*}{\partial a_{xy}}$ is non-negative. 
\end{proof}

\section{Proof of Proposition \ref{proposition:completegraph}}
\begin{proof}
When ${\mathcal G}_{\rm u}$ is a complete graph and has homogeneous edge weights, there is a resistor between any two nodes (i.e., locations) in the electrical network. Moreover, these resistors have the same resistance. In this case, the effective resistances between different locations are the same. Hence, if arcs $\left(i,j\right)$ and $\left(x,y\right)$ do not have a common vertex, Eq. (\ref{equ:impactadprice}) becomes
\begin{align}
\frac{\partial p_{ij}^*}{\partial a_{xy}}=\frac{\theta_{xy}}{4\xi_{ij}}\left(R_{jx}-R_{ix}-R_{jy}+R_{iy}\right)=0,
\end{align}
where the second equality is based on $R_{jx}=R_{ix}=R_{jy}=R_{iy}$. We can see that $p_{ij}^*$ does not change with $a_{xy}$.
\end{proof}

\section{Proof of Proposition \ref{proposition:cutvertex}}
\begin{proof}
Recall that the effective resistance between two locations is defined as the voltage between them when a unit current is injected at one location and withdrawn at the other location. Since ${\mathcal N}_1$ and ${\mathcal N}_2$ are the vertex sets of two connected components of ${\mathcal G}_{\rm u}-k$, if $i\in{\mathcal N}_1 \cup \left\{k\right\}$ and $x\in{\mathcal N}_2 \cup \left\{k\right\}$, vertex $k$ is on all paths between $i$ and $x$. Then, the effective resistances between $i$, $x$, and $k$ satisfy $R_{ix}=R_{ik}+R_{kx}$. Similarly, for $i,j\in{\mathcal N}_1 \cup \left\{k\right\}$ and $x,y\in{\mathcal N}_2 \cup \left\{k\right\}$, we have $R_{jx}=R_{jk}+R_{kx}$, $R_{jy}=R_{jk}+R_{ky}$, and $R_{iy}=R_{ik}+R_{ky}$. Based on these four equations, we can see that $R_{jx}-R_{ix}$ and $R_{jy}-R_{iy}$ are equal (i.e., $R_{jx}-R_{ix}=R_{jy}-R_{iy}= R_{jk}-R_{ik}$).

According to (\ref{equ:impactadprice}), $\frac{\partial p_{ij}^*}{\partial a_{xy}}$ is given by
\begin{align}
\frac{\partial p_{ij}^*}{\partial a_{xy}} =-\frac{1}{2}{\mathbbm 1}_{\left\{i=x,j=y\right\}}+\frac{\theta_{xy}}{4\xi_{ij}}\left(R_{jx}-R_{ix}-R_{jy}+R_{iy}\right). 
\end{align}
For $i,j\in{\mathcal N}_1 \cup \left\{k\right\}$ and $x,y\in{\mathcal N}_2 \cup \left\{k\right\}$, the relations $i=x$ and $j=y$ cannot hold in the same time. This means that ${\mathbbm 1}_{\left\{i=x,j=y\right\}}=0$. As analyzed above, $R_{jx}-R_{ix}$ and $R_{jy}-R_{iy}$ are equal. Therefore, the value of $\frac{\partial p_{ij}^*}{\partial a_{xy}}$ is zero, which implies that $p_{ij}^*$ does not change with $a_{xy}$.
\end{proof}

\section{Proofs of Propositions \ref{proposition:platpayoff} and \ref{proposition:optimalCS}}
\begin{proof}
First, as shown in (\ref{opt2:objective}), the provider's payoff equals $\sum_{\left(i,j\right)\in{\mathcal A}} \theta_{ij}\xi_{ij} \left(1-p_{ij}\right)  \left(p_{ij}+a_{ij}-c\right)$. Its optimal pricing is given by $p_{ij}^*=\frac{1-a_{ij}+c}{2}+\frac{1}{4\xi_{ij}} \sum_{k\in{\mathcal N}} \left(R_{jk}-R_{ik}\right) v_k$. Hence, the provider's payoff under the optimal pricing is 
\begin{align}
\Pi^{\rm provider}= \sum_{\left(i,j\right)\in {\mathcal A}} \theta_{ij}\xi_{ij} \left(\left(\frac{1+a_{ij}-c}{2}\right)^2 - \left(\frac{\sum_{k\in{\mathcal N}} \left(R_{jk}-R_{ik}\right) v_k}{4 \xi_{ij}}\right)^2 \right).
\end{align}

Second, as discussed in Section \ref{sec:cs}, the total consumer surplus equals $\sum_{\left(i,j\right)\in{\mathcal A}} \frac{1}{2} \theta_{ij} \xi_{ij} \left(1-p_{ij}\right)^2$. By substituting the expression of $p_{ij}^*$ into it, we have
\begin{align}
{\rm CS}=\sum_{\left(i,j\right)\in{\mathcal A}} \frac{1}{2} \theta_{ij} \xi_{ij} \left(\frac{1+a_{ij}-c}{2}-\frac{\sum_{k\in{\mathcal N}} \left(R_{jk}-R_{ik}\right) v_k}{4\xi_{ij} }\right)^2.
\end{align}
\end{proof}

\section{Proof of Theorem \ref{theorem:payoff2CS}}\label{appendix:sec:payoff2CS}
\begin{proof}
First, to simplify the presentation, we represent $\Pi^{\rm provider}$ and ${\rm CS}$ using the optimal dual variables ${\bm \lambda}^*$ as follows:
\begin{align}
& \Pi^{\rm provider}= \sum_{\left(i,j\right)\in{\mathcal A}} \theta_{ij}\xi_{ij} \left(\frac{1+a_{ij}-c}{2}-\frac{\lambda_i^*-\lambda_j^*}{2\xi_{ij}}\right)\left(\frac{1+a_{ij}-c}{2}+\frac{\lambda_i^*-\lambda_j^*}{2\xi_{ij}}\right),\\
& {\rm CS}=\sum_{\left(i,j\right)\in{\mathcal A}} \frac{1}{2} \theta_{ij} \xi_{ij} \left(\frac{1+a_{ij}-c}{2}-\frac{\lambda_i^*-\lambda_j^*}{2\xi_{ij}}\right)^2.
\end{align}

Second, we compute $\Pi^{\rm provider}-2{\rm CS}$, and the detailed steps are shown as follows.
\begin{align}
\nonumber
& \Pi^{\rm provider}-2{\rm CS} \\
\nonumber
=&\sum_{\left(i,j\right)\in{\mathcal A}} \theta_{ij} \left(\frac{1+a_{ij}-c}{2}-\frac{\lambda_i^*-\lambda_j^*}{2\xi_{ij}}\right)\left(\lambda_i^*-\lambda_j^*\right) \\
\nonumber
=& \sum_{\left(i,j\right)\in{\mathcal A}} \theta_{ij} \left(\frac{1+a_{ij}-c}{2}-\frac{\lambda_i^*-\lambda_j^*}{2\xi_{ij}}\right) \lambda_i^* - \sum_{\left(i,j\right)\in{\mathcal A}} \theta_{ij} \left(\frac{1+a_{ij}-c}{2}-\frac{\lambda_i^*-\lambda_j^*}{2\xi_{ij}}\right) \lambda_j^* \\
\nonumber
\overset{(a)}= & \sum_{i\in{\mathcal N}} \sum_{j\in{\mathcal N}} \theta_{ij} \left(\frac{1+a_{ij}-c}{2}-\frac{\lambda_i^*-\lambda_j^*}{2\xi_{ij}}\right) \lambda_i^*  - \sum_{i\in{\mathcal N}} \sum_{j\in{\mathcal N}} \theta_{ij} \left(\frac{1+a_{ij}-c}{2}-\frac{\lambda_i^*-\lambda_j^*}{2\xi_{ij}}\right) \lambda_j^* \\
\nonumber
 = & \sum_{i\in{\mathcal N}} \sum_{j\in{\mathcal N}} \theta_{ij} \left(\frac{1+a_{ij}-c}{2}-\frac{\lambda_i^*-\lambda_j^*}{2\xi_{ij}}\right) \lambda_i^*  - \sum_{i\in{\mathcal N}} \sum_{j\in{\mathcal N}} \theta_{ji} \left(\frac{1+a_{ji}-c}{2}-\frac{\lambda_j^*-\lambda_i^*}{2\xi_{ji}}\right) \lambda_i^* \\
 \nonumber
= & \frac{1}{2}\sum_{i\in{\mathcal N}} \lambda_i^* \sum_{j\in{\mathcal N}} \left(\theta_{ij} \left(1+a_{ij}-c\right) - \theta_{ji} \left(1+a_{ji}-c\right)   \right) -\frac{1}{2} \sum_{i\in{\mathcal N}} \lambda_i^* \sum_{j\in{\mathcal N}} \left( \frac{\theta_{ij}}{\xi_{ij}}+ \frac{\theta_{ji}}{\xi_{ji}}\right) \left(\lambda_i^*-\lambda_j^*\right) \\
\nonumber
\overset{(b)}= & \frac{1}{2}\sum_{i\in{\mathcal N}} \lambda_i^* v_i -\frac{1}{2} \sum_{i\in{\mathcal N}} \lambda_i^* \sum_{j\in{\mathcal N}} \left( \frac{\theta_{ij}}{\xi_{ij}}+ \frac{\theta_{ji}}{\xi_{ji}}\right) \left(\lambda_i^*-\lambda_j^*\right) = \frac{1}{2}\sum_{i\in{\mathcal N}} \lambda_i^* \left(v_i - \sum_{j\in{\mathcal N}} \left( \frac{\theta_{ij}}{\xi_{ij}}+ \frac{\theta_{ji}}{\xi_{ji}}\right) \left(\lambda_i^*-\lambda_j^*\right)  \right) \\
\overset{(c)}=& \frac{1}{2}\sum_{i\in{\mathcal N}} \lambda_i^* \left(v_i - \sum_{j\in{\mathcal N}} \left( \frac{\theta_{ij}}{\xi_{ij}}+ \frac{\theta_{ji}}{\xi_{ji}}\right) \left( \frac{1}{2}\sum_{k\in{\mathcal N}} \left(R_{jk}-R_{ik}\right)v_k \right)  \right).
\label{appendix:long:CSpayoff}
\end{align}
We use the fact that $\theta_{ij}=0$ for all $\left(i,j\right)\notin {\mathcal A}$ to derive equality (a). We again use this fact and the definition of $v_i$ in (\ref{equ:valueoflocation}) to derive equality (b). When deriving equality (c), we use the result of $\lambda_i^*-\lambda_j^*$ in (\ref{equ:difflambda}).

Third, we analyze $\sum_{j\in{\mathcal N}} \left( \frac{\theta_{ij}}{\xi_{ij}}+ \frac{\theta_{ji}}{\xi_{ji}}\right) \left( \frac{1}{2}\sum_{k\in{\mathcal N}} \left(R_{jk}-R_{ik}\right)v_k \right)$, which is part of the result in (\ref{appendix:long:CSpayoff}). 
According to the local sum rules of the resistances \cite{chen2010random} (shown in (\ref{equ:localsum}) in our paper), we have 
\begin{align}
\sum_{j:\left(i,j\right)\in{\mathcal E}} \frac{R_{ij}+R_{ik}-R_{jk}}{r_{ij}} =2, \forall i\ne k, i,k\in{\mathcal N}.
\end{align}
Using $r_{ij}=\frac{1}{\frac{\theta_{ij}}{\xi_{ij}}+\frac{\theta_{ji}}{\xi_{ji}}}$ and the fact that $\theta_{ij}=\theta_{ji}=0$ when $\left(i,j\right)\notin {\mathcal E}$, we can rewrite the above equation as follows:
\begin{align}
\sum_{j\in{\mathcal N}} \left(\frac{\theta_{ij}}{\xi_{ij}}+\frac{\theta_{ji}}{\xi_{ji}}\right) \left(R_{ij}+R_{ik}-R_{jk}\right) =2, \forall i\ne k, i,k\in{\mathcal N}.
\end{align}
By rearranging the above equation, we get the following relation:
\begin{align}
\sum_{j\in{\mathcal N}} \left(\frac{\theta_{ij}}{\xi_{ij}}+\frac{\theta_{ji}}{\xi_{ji}}\right) \left(R_{jk}-R_{ik}\right)=\left(\sum_{j\in{\mathcal N}} \left(\frac{\theta_{ij}}{\xi_{ij}}+\frac{\theta_{ji}}{\xi_{ji}}\right) R_{ij}\right)-2, \forall i\ne k, i,k\in{\mathcal N}.\label{appendix:equ:localsum}
\end{align}
Next, we rearrange $\sum_{j\in{\mathcal N}} \left( \frac{\theta_{ij}}{\xi_{ij}}+ \frac{\theta_{ji}}{\xi_{ji}}\right) \left( \frac{1}{2}\sum_{k\in{\mathcal N}} \left(R_{jk}-R_{ik}\right)v_k \right)$ and show the steps as follows.

\begin{align}
\nonumber
& \sum_{j\in{\mathcal N}} \left( \frac{\theta_{ij}}{\xi_{ij}}+ \frac{\theta_{ji}}{\xi_{ji}}\right) \left( \frac{1}{2}\sum_{k\in{\mathcal N}} \left(R_{jk}-R_{ik}\right)v_k \right) \\
\nonumber
= &\frac{1}{2}\sum_{k\in{\mathcal N}}\sum_{j\in{\mathcal N}} \left( \frac{\theta_{ij}}{\xi_{ij}}+ \frac{\theta_{ji}}{\xi_{ji}}\right) \left(R_{jk}-R_{ik}\right)v_k\\
\nonumber
=& \frac{1}{2}\sum_{k\ne i, k\in{\mathcal N}} \left(v_k \sum_{j\in{\mathcal N}} \left( \frac{\theta_{ij}}{\xi_{ij}}+ \frac{\theta_{ji}}{\xi_{ji}}\right) \left(R_{jk}-R_{ik}\right)\right)  + \frac{1}{2} \sum_{j\in{\mathcal N}} \left( \frac{\theta_{ij}}{\xi_{ij}}+ \frac{\theta_{ji}}{\xi_{ji}}\right) \left(R_{ji}-R_{ii}\right)v_i \\
\nonumber
\overset{(a)}=& \frac{1}{2}\sum_{k\ne i, k\in{\mathcal N}}\left(\left(\sum_{j\in{\mathcal N}} \left(\frac{\theta_{ij}}{\xi_{ij}}+\frac{\theta_{ji}}{\xi_{ji}}\right) R_{ij}\right)-2\right) v_k  + \frac{1}{2} \sum_{j\in{\mathcal N}} \left( \frac{\theta_{ij}}{\xi_{ij}}+ \frac{\theta_{ji}}{\xi_{ji}}\right) R_{ji} v_i\\
\nonumber
=& \frac{1}{2}\sum_{k\ne i, k\in{\mathcal N}}\left(\sum_{j\in{\mathcal N}} \left(\frac{\theta_{ij}}{\xi_{ij}}+\frac{\theta_{ji}}{\xi_{ji}}\right) R_{ij}\right) v_k  + \frac{1}{2} \sum_{j\in{\mathcal N}} \left( \frac{\theta_{ij}}{\xi_{ij}}+ \frac{\theta_{ji}}{\xi_{ji}}\right) R_{ji} v_i - \sum_{k\ne i, k\in{\mathcal N}} v_k \\
\nonumber
=& \frac{1}{2} \sum_{j\in{\mathcal N}} \sum_{k\ne i, k\in{\mathcal N}}  \left(\frac{\theta_{ij}}{\xi_{ij}}+\frac{\theta_{ji}}{\xi_{ji}}\right) R_{ij}  v_k  + \frac{1}{2} \sum_{j\in{\mathcal N}} \left( \frac{\theta_{ij}}{\xi_{ij}}+ \frac{\theta_{ji}}{\xi_{ji}}\right) R_{ji} v_i - \sum_{k\ne i, k\in{\mathcal N}} v_k \\
\nonumber
\overset{(b)}=& \frac{1}{2} \sum_{j\in{\mathcal N}} \sum_{k\ne i, k\in{\mathcal N}}  \left(\frac{\theta_{ij}}{\xi_{ij}}+\frac{\theta_{ji}}{\xi_{ji}}\right) R_{ij}  v_k  + \frac{1}{2} \sum_{j\in{\mathcal N}} \left( \frac{\theta_{ij}}{\xi_{ij}}+ \frac{\theta_{ji}}{\xi_{ji}}\right) R_{ij} v_i - \sum_{k\ne i, k\in{\mathcal N}} v_k \\
\nonumber
=& \frac{1}{2} \sum_{j\in{\mathcal N}} \left(\frac{\theta_{ij}}{\xi_{ij}}+\frac{\theta_{ji}}{\xi_{ji}}\right) R_{ij}  \left(\left(\sum_{k\ne i, k\in{\mathcal N}}    v_k\right)+v_i\right) - \sum_{k\ne i, k\in{\mathcal N}} v_k\\
\overset{(c)} =& v_i.\label{appendix:equ:long:usinglocal}
\end{align}

We can see that $\sum_{j\in{\mathcal N}} \left( \frac{\theta_{ij}}{\xi_{ij}}+ \frac{\theta_{ji}}{\xi_{ji}}\right) \left( \frac{1}{2}\sum_{k\in{\mathcal N}} \left(R_{jk}-R_{ik}\right)v_k \right)$ equals $v_i$. Specifically, we use (\ref{appendix:equ:localsum}) and $R_{ii}=0$ to get equality (a), and use $R_{ji}=R_{ij}$ to get equality (b). When deriving equality (c), we use the fact that $\sum_{k\in{\mathcal N}} v_k=0$. Combining (\ref{appendix:long:CSpayoff}) and (\ref{appendix:equ:long:usinglocal}), we can prove that $\Pi^{\rm provider}-2{\rm CS}=0$.

Fourth, from the provider's problem (\ref{opt:sumsub:pricing}), we can see that given any pricing, the provider's payoff is non-decreasing in $a_{xy}$ for any $\left(x,y\right)\in{\mathcal A}$. Hence, the provider's optimal payoff is non-decreasing in $a_{xy}$. Because $\Pi^{\rm provider}=2{\rm CS}$, we can see that ${\rm CS}$ is also non-decreasing in $a_{xy}$.
\end{proof}

\section{Proof of Proposition \ref{proposition:deltaa}}
\begin{proof}
There are two approaches to derive $\Delta\left({\bm a}\right)$ using the local sum rules. Here, we introduce the first approach. The first step is to prove $\Pi^{\rm provider}=2{\rm CS}$ based on the local sum rules. The details of the proof are given in Section \ref{appendix:sec:payoff2CS}. Then, we have the following two (equivalent) expressions for $\Pi^{\rm provider}$:
\begin{align}
&\Pi^{\rm provider}= \sum_{\left(i,j\right)\in {\mathcal A}} \theta_{ij}\xi_{ij} \left(\left(\frac{1+a_{ij}-c}{2}\right)^2 - \left(\frac{\sum_{k\in{\mathcal N}} \left(R_{jk}-R_{ik}\right) v_k}{4 \xi_{ij}}\right)^2 \right),\label{appendix:equ:payofflong1}\\
&\Pi^{\rm provider}=\sum_{\left(i,j\right)\in{\mathcal A}} \theta_{ij} \xi_{ij} \left(\frac{1+a_{ij}-c}{2}-\frac{\sum_{k\in{\mathcal N}} \left(R_{jk}-R_{ik}\right) v_k}{4\xi_{ij} }\right)^2.\label{appendix:equ:payofflong2}
\end{align}
The first expression is from (\ref{equ:opprice:payoff}), and the second expression is from $\Pi^{\rm provider}=2{\rm CS}$ and (\ref{equ:opprice:cs}). 

The second step is to rearrange $\sum_{\left(i,j\right)\in {\mathcal A}} \theta_{ij}\xi_{ij} \left(\frac{\sum_{k\in{\mathcal N}} \left(R_{jk}-R_{ik}\right) v_k}{4 \xi_{ij}}\right)^2 $. Because the right sides of (\ref{appendix:equ:payofflong1}) and (\ref{appendix:equ:payofflong2}) are equal, we can derive the following results: 

\begin{align}
\nonumber
& \sum_{\left(i,j\right)\in {\mathcal A}} \theta_{ij}\xi_{ij} \left(\left(\frac{1+a_{ij}-c}{2}\right)^2 - \left(\frac{\sum_{k\in{\mathcal N}} \left(R_{jk}-R_{ik}\right) v_k}{4 \xi_{ij}}\right)^2 \right) \\
\nonumber
& {~~}=\sum_{\left(i,j\right)\in{\mathcal A}} \theta_{ij} \xi_{ij} \left(\frac{1+a_{ij}-c}{2}-\frac{\sum_{k\in{\mathcal N}} \left(R_{jk}-R_{ik}\right) v_k}{4\xi_{ij} }\right)^2 \\
\nonumber
 \Leftrightarrow & \sum_{\left(i,j\right)\in {\mathcal A}} \theta_{ij}\xi_{ij} \left(\frac{1+a_{ij}-c}{2}\right)^2 - \sum_{\left(i,j\right)\in {\mathcal A}} \theta_{ij}\xi_{ij}  \left(\frac{\sum_{k\in{\mathcal N}} \left(R_{jk}-R_{ik}\right) v_k}{4 \xi_{ij}}\right)^2 
\\
\nonumber
& {~~}= \sum_{\left(i,j\right)\in{\mathcal A}} \theta_{ij} \xi_{ij} \left(\frac{1+a_{ij}-c}{2}\right)^2 + \sum_{\left(i,j\right)\in{\mathcal A}} \theta_{ij} \xi_{ij} \left(\frac{\sum_{k\in{\mathcal N}} \left(R_{jk}-R_{ik}\right) v_k}{4\xi_{ij} }\right)^2 \\
\nonumber
& {~~}{~~}{~~}{~~}{~~}{~~} -\sum_{\left(i,j\right)\in{\mathcal A}} \theta_{ij} \xi_{ij} \left(1+a_{ij}-c\right)\left(\frac{\sum_{k\in{\mathcal N}} \left(R_{jk}-R_{ik}\right) v_k}{4\xi_{ij} }\right)\\
 \Leftrightarrow & \sum_{\left(i,j\right)\in{\mathcal A}} \theta_{ij} \xi_{ij} \left(\frac{\sum_{k\in{\mathcal N}} \left(R_{jk}-R_{ik}\right) v_k}{4\xi_{ij} }\right)^2 =  \sum_{\left(i,j\right)\in{\mathcal A}} \frac{1}{8} \theta_{ij} \left(1+a_{ij}-c\right)\left(\sum_{k\in{\mathcal N}} \left(R_{jk}-R_{ik}\right) v_k\right).\label{appendix:equ:equivalence}
\end{align}

Hence, $\sum_{\left(i,j\right)\in{\mathcal A}}\! \theta_{ij} \xi_{ij} \left(\frac{\sum_{k\in{\mathcal N}} \left(R_{jk}-R_{ik}\right) v_k}{4\xi_{ij} }\right)^2$ equals $\sum_{\left(i,j\right)\in{\mathcal A}} \frac{1}{8} \theta_{ij} \left(1+a_{ij}-c\right)\!\left(\sum_{k\in{\mathcal N}} \left(R_{jk}-R_{ik}\right) v_k\right)$. Then, we can replace the term $\sum_{\left(i,j\right)\in{\mathcal A}} \theta_{ij} \xi_{ij} \left(\frac{\sum_{k\in{\mathcal N}} \left(R_{jk}-R_{ik}\right) v_k}{4\xi_{ij} }\right)^2$ in the optimal payoff in (\ref{appendix:equ:payofflong1}) by $\sum_{\left(i,j\right)\in{\mathcal A}} \frac{1}{8} \theta_{ij} \left(1+a_{ij}-c\right)\left(\sum_{k\in{\mathcal N}} \left(R_{jk}-R_{ik}\right) v_k\right)$, and get the following result:
\begin{align}
\Pi^{\rm provider}=\sum_{\left(i,j\right)\in{\mathcal A}} \theta_{ij} \xi_{ij} \left(\frac{1+a_{ij}-c}{2}\right)^2 -\sum_{\left(i,j\right)\in{\mathcal A}} \left(\frac{1}{8}\theta_{ij} \left(1+a_{ij}-c\right)\sum_{k\in{\mathcal N}} \left(R_{jk}-R_{ik}\right)v_k\right),
\end{align}
where the right side is $\Delta\left({\bm a}\right)$.
\end{proof}

\section{Proof of Theorem \ref{theorem:arcbad}}\label{appendix:sec:arcbad}
\begin{proof}
When the provider collaborates with the advertiser corresponding to arc $\left(x,y\right)$, it only has a positive unit ad revenue on arc $\left(x,y\right)$, i.e., $a_{xy}=b_{xy}$ and $a_{ij}=0$ for other arcs. Based on this and the condition that $\theta_{ij}=\theta_{ji}$ for all $\left(i,j\right)\in{\mathcal A}$, we can compute the provider's payoff under the collaboration choice $\left(x,y\right)$ as follows.

\begin{align}
\nonumber
& \Pi^{\rm provider} \\
\nonumber
=& \sum_{\left(i,j\right)\in{\mathcal A}} \theta_{ij} \xi_{ij} \left(\frac{1+a_{ij}-c}{2}\right)^2 -\sum_{\left(i,j\right)\in{\mathcal A}} \left(\frac{1}{8}\theta_{ij} \left(1+a_{ij}-c\right)\sum_{k\in{\mathcal N}} \left(R_{jk}-R_{ik}\right)v_k\right) \\
\nonumber 
=& \sum_{\left(i,j\right)\ne \left(x,y\right),\left(i,j\right)\in{\mathcal A}} \theta_{ij} \xi_{ij} \left(\frac{1-c}{2}\right)^2 + \theta_{xy} \xi_{xy} \left(\frac{1+b_{xy}-c}{2}\right)^2 \\
\nonumber
& -\!\!\!\sum_{\left(i,j\right)\ne \left(x,y\right),\left(i,j\right)\in{\mathcal A}} \left(\frac{1}{8}\theta_{ij} \left(1-c\right)\sum_{k\in{\mathcal N}} \left(R_{jk}-R_{ik}\right)v_k\right)
-\frac{1}{8}\theta_{xy} \left(1+b_{xy}-c\right)\sum_{k\in{\mathcal N}} \left(R_{yk}-R_{xk}\right)v_k\\
\nonumber 
=& \sum_{\left(i,j\right)\in{\mathcal A}} \theta_{ij} \xi_{ij} \left(\frac{1-c}{2}\right)^2 + \theta_{xy} \xi_{xy} \left(\frac{b_{xy}}{2}\right)^2 + \theta_{xy} \xi_{xy} \frac{b_{xy}}{2} \left(1-c\right) \\
\nonumber
& -\sum_{\left(i,j\right)\in{\mathcal A}} \left(\frac{1}{8}\theta_{ij} \left(1-c\right)\sum_{k\in{\mathcal N}} \left(R_{jk}-R_{ik}\right)v_k\right)
-\frac{1}{8}\theta_{xy} b_{xy}\sum_{k\in{\mathcal N}} \left(R_{yk}-R_{xk}\right)v_k \\
\nonumber
\overset{(a)}=& \sum_{\left(i,j\right)\in{\mathcal A}} \theta_{ij} \xi_{ij} \left(\frac{1-c}{2}\right)^2 + \theta_{xy} \xi_{xy} \left(\frac{b_{xy}}{2}\right)^2 + \theta_{xy} \xi_{xy} \frac{b_{xy}}{2} \left(1-c\right)-\frac{1}{8}\theta_{xy} b_{xy}\sum_{k\in{\mathcal N}} \left(R_{yk}-R_{xk}\right)v_k \\
\nonumber 
\overset{(b)}=& \sum_{\left(i,j\right)\in{\mathcal A}} \theta_{ij} \xi_{ij} \left(\frac{1-c}{2}\right)^2 + \theta_{xy} \xi_{xy} \left(\frac{b_{xy}}{2}\right)^2 + \theta_{xy} \xi_{xy} \frac{b_{xy}}{2} \left(1-c\right) \\
\nonumber
&{~~}{~~}{~~} -\frac{1}{8}\theta_{xy} b_{xy} \left(\left(R_{yy}-R_{xy}\right)\left(-\theta_{xy}b_{xy}\right)+\left(R_{yx}-R_{xx}\right)\theta_{xy}b_{xy}\right) \\
\overset{(c)}=& \sum_{\left(i,j\right)\in{\mathcal A}} \theta_{ij} \xi_{ij} \left(\frac{1-c}{2}\right)^2 + \theta_{xy} \xi_{xy} \left(\frac{b_{xy}}{2}\right)^2 + \theta_{xy} \xi_{xy} \frac{b_{xy}}{2} \left(1-c\right) -\frac{1}{4}\theta_{xy}^2 b_{xy}^2 R_{xy}.\label{appendix:equ:locationad:theorem}
\end{align} 

Note that when $\theta_{ij}=\theta_{ji}$ for all $\left(i,j\right)\in{\mathcal A}$, we have the following relation:
\begin{align}
\frac{1}{8}\theta_{ji} \left(1-c\right)\sum_{k\in{\mathcal N}} \left(R_{ik}-R_{jk}\right)v_k\!=\! \frac{1}{8}\theta_{ij} \left(1-c\right)\sum_{k\in{\mathcal N}} \left(R_{ik}-R_{jk}\right)v_k 
\!=\! -\frac{1}{8}\theta_{ij} \left(1-c\right)\sum_{k\in{\mathcal N}} \left(R_{jk}-R_{ik}\right)v_k.
\end{align}
This implies that the sum of $\frac{1}{8}\theta_{ji} \left(1-c\right)\sum_{k\in{\mathcal N}} \left(R_{ik}-R_{jk}\right)v_k$ and $\frac{1}{8}\theta_{ij} \left(1-c\right)\sum_{k\in{\mathcal N}} \left(R_{jk}-R_{ik}\right)v_k$ is zero. If $\left(i,j\right)\in{\mathcal A}$, since $\theta_{ij}=\theta_{ji}$ for all $\left(i,j\right)\in{\mathcal A}$, we have $\left(j,i\right)\in{\mathcal A}$. Therefore, we can derive equality (a) in (\ref{appendix:equ:locationad:theorem}). According to the definition of $v_i$ in (\ref{equ:valueoflocation}), when $\theta_{ij}=\theta_{ji}$ for all $\left(i,j\right)\in{\mathcal A}$ and the provider only collaborates with $\left(x,y\right)$, we can see that $v_x=\theta_{xy}b_{xy}$, $v_y=-\theta_{xy}b_{xy}$, and $v_i=0$ for $i\ne x,y$. This enables us to derive equality (b). When deriving equality (c), we use the property that $R_{yy}=0$, $R_{xx}=0$, and $R_{yx}=R_{xy}$.

The provider needs to select $\left(x,y\right)$ to maximize its payoff in (\ref{appendix:equ:locationad:theorem}). We can see that the term $\sum_{\left(i,j\right)\in{\mathcal A}} \theta_{ij} \xi_{ij} \left(\frac{1-c}{2}\right)^2$ does not change with the provider's choice of $\left(x,y\right)$. Hence, the provider should select $\left(x,y\right)$ to maximize the following expression:
\begin{align}
\frac{1}{4}\theta_{xy} \xi_{xy} b_{xy}^2 + \frac{1}{2} \theta_{xy} \xi_{xy} b_{xy} \left(1-c\right) -\frac{1}{4}\theta_{xy}^2 b_{xy}^2 R_{xy}.\label{appendix:equ:tobemax}
\end{align}
We can multiply it by $4$, which corresponds to the expression to be maximized in Theorem \ref{theorem:arcbad}.

At last, we prove that (\ref{appendix:equ:tobemax}) increases with $b_{xy}$ for all $\left(x,y\right)\in{\mathcal A}$. In (\ref{appendix:equ:tobemax}), the coefficient of $b_{xy}^2$ is $\frac{1}{4}\theta_{xy} \left(\xi_{xy} -\theta_{xy} R_{xy}\right)$. The effective resistance between two locations is always no greater than the resistance of the resistor between them. Hence, $R_{xy}\le r_{xy}=\frac{1}{\frac{\theta_{xy}}{\xi_{xy}}+\frac{\theta_{yx}}{\xi_{yx}}}\le \frac{\xi_{xy}}{\theta_{xy}}$. This means that the coefficient of $b_{xy}^2$ in (\ref{appendix:equ:tobemax}) is non-negative. Moreover, since $c<1$ (i.e., the assumption of $c$ in the model), the coefficient of $b_{xy}$ in (\ref{appendix:equ:tobemax}) is positive. Therefore, (\ref{appendix:equ:tobemax}) increases with $b_{xy}$ for all $\left(x,y\right)\in{\mathcal A}$.
\end{proof}

\section{Proof of Corollary \ref{corollary:arcbad}}
\begin{proof}
If $\theta_{ij}=\theta_{ji}={\bar \theta}$, $\xi_{ij}={\bar \xi}$, and $b_{ij}={\bar b}$ for all $\left(i,j\right)\in{\mathcal A}$, (\ref{equ:theorem:arcbad}) becomes
\begin{align}
\left(x,y\right)={\argmax_{\left(i,j\right)\in{\mathcal A}}} {~}{\bar \theta}{\bar \xi} \left({\bar b}^2+2\left(1-c\right){\bar b}\right)- {\bar \theta}^2 {\bar b}^2 R_{ij}.
\end{align}
The term ${\bar \theta}{\bar \xi} \left({\bar b}^2+2\left(1-c\right){\bar b}\right)$ is independent of the choice $\left(x,y\right)$. We can see that the above condition for choosing $\left(x,y\right)$ is equivalent to $\left(x,y\right)={\argmin_{\left(i,j\right)\in{\mathcal A}}} R_{ij}$.
\end{proof}

\section{Optimality of Search Steps in Section \ref{subsubsec:arcbad:general}}\label{appendix:sec:algorithm}
We show that the steps which we introduce to search for the collaboration choice $\left(x,y\right)$ lead to the optimal collaboration choice. We only need to prove that when $h<\left| {\mathcal A} \right|$, the provider's payoff when it collaborates with the advertiser corresponding to the $h$-th arc is no less than its payoff when it collaborates with the advertiser corresponding to the $i$-th arc ($i>h$). Recall that the arcs are sorted in their descending order of their corresponding $\Delta\left({\bm a}\right)$.

To simplify the presentation, we introduce some new notations. We use ${\tilde l}_{i}$ ($1 \le i \le \left| {\mathcal A} \right|$) to denote the arc with the $i$-th largest $\Delta\left({\bm a}\right)$. Moreover, we use ${\bm a}_{{\tilde l}_i}$ to denote the unit ad revenue vector when the provider collaborates with the advertiser corresponding to ${\tilde l}_{i}$. Hence, we have the following relation:
\begin{align}
\Delta\left({\bm a}_{{\tilde l}_h}\right) \ge \Delta\left({\bm a}_{{\tilde l}_{h+1}}\right) \ge \ldots \ge \Delta\left({\bm a}_{{\tilde l}_{\left|{\mathcal A}\right|}}\right).\label{appendix:equ:compare}
\end{align}
Furthermore, we have $\Pi^{\rm provider} \left( {\bm a}_{{\tilde l}_i}\right)\le \Delta\left({\bm a}_{{\tilde l}_i}\right)$ for all arcs ${\tilde l}_i$. This is because $\Pi^{\rm provider} \left( {\bm a}_{{\tilde l}_i}\right)$ denotes the provider's optimal payoff with constraint $p_{ij}\le 1$ (for all $\left(i,j\right)\in{\mathcal A}$), and the expression of $ \Delta\left({\bm a}_{{\tilde l}_i}\right)$ equals the provider's optimal payoff when there is no constraint $p_{ij}\le 1$. In particular, when $\mu_{ij}^*=0$ for all $\left(i,j\right)\in{\mathcal A}$, the equality holds, i.e., $\Pi^{\rm provider} \left( {\bm a}_{{\tilde l}_i}\right)= \Delta\left({\bm a}_{{\tilde l}_i}\right)$. 

When $h<\left| {\mathcal A} \right|$, if the provider collaborates with the advertiser corresponding to arc ${\tilde l}_h$, we have $\mu_{ij}^*=0$ for all $\left(i,j\right)\in{\mathcal A}$ (based on the definition of $h$). This means that $\Pi^{\rm provider} \left( {\bm a}_{{\tilde l}_h}\right)= \Delta\left({\bm a}_{{\tilde l}_h}\right)$. Considering (\ref{appendix:equ:compare}), we have the following relation:
\begin{align}
\Pi^{\rm provider} \left( {\bm a}_{{\tilde l}_h}\right) \ge \Delta\left({\bm a}_{{\tilde l}_{h+1}}\right) \ge \ldots \ge \Delta\left({\bm a}_{{\tilde l}_{\left|{\mathcal A}\right|}}\right).
\end{align}
Because $\Pi^{\rm provider} \left( {\bm a}_{{\tilde l}_i}\right)\le \Delta\left({\bm a}_{{\tilde l}_i}\right)$ for all arcs ${\tilde l}_i$, we can further derive the following result:
\begin{align}
\Pi^{\rm provider} \left( {\bm a}_{{\tilde l}_h}\right)\ge \Pi^{\rm provider} \left( {\bm a}_{{\tilde l}_i}\right), \forall i>h.
\end{align}
This means that the provider's payoff when it collaborates with the advertiser corresponding to the $h$-th arc is no less than its payoff when it collaborates with the advertiser corresponding to the $i$-th arc ($i>h$). As a result, the provider only needs to compare the first $h$ arcs to determine the optimal arc, as described in \emph{Step 3} in Section \ref{subsubsec:arcbad:general}.

\section{Proof of Theorem \ref{theorem:locationad}}
\begin{proof}
When the provider collaborates with advertiser $k$, it only has positive ad revenues on the arcs pointing to $k$. Based on this and the condition that $\theta_{ij}=\theta_{ji}$ for all $\left(i,j\right)\in{\mathcal A}$, we can compute the provider's payoff under the collaboration choice $k$ in (\ref{appendix:equ:lastlongeqs}). 

Note that we use $\left(i,j\right)\in{\mathcal A}:j=k$ to denote an arc $\left(i,j\right)$ that belongs to set ${\mathcal A}$ and satisfies $j=k$. Similarly, we use $n\in{\mathcal N}:n\ne k, \left(n,k\right)\notin{\mathcal A}$ to denote a location $n$ that belongs to set $\mathcal N$, is not $k$, and has no arc pointing to $k$. When deriving equality (a), we use the relation that $\sum_{\left(i,j\right)\in{\mathcal A}} \left(\frac{1}{8}\theta_{ij} \left(1-c\right)\sum_{n\in{\mathcal N}} \left(R_{jn}-R_{in}\right)v_n\right)=0$. We have proved this in Section \ref{appendix:sec:arcbad}. The basic idea is that when $\theta_{ij}=\theta_{ji}$ for all $\left(i,j\right)\in{\mathcal A}$, the sum of $\frac{1}{8}\theta_{ij} \left(1-c\right)\sum_{n\in{\mathcal N}} \left(R_{jn}-R_{in}\right)v_n$ and $\frac{1}{8}\theta_{ji} \left(1-c\right)\sum_{n\in{\mathcal N}} \left(R_{in}-R_{jn}\right)v_n$ is zero. Moreover, if $\left(i,j\right)\in{\mathcal A}$, then $\left(j,i\right)\in{\mathcal A}$. When deriving equality (b), we use the expressions of different $v_n$. When $\theta_{ij}=\theta_{ji}$ for all $\left(i,j\right)\in{\mathcal A}$ and the provider only collaborates with the location-based advertiser $k$, we have $v_k=-\sum_{n\in{\mathcal N}:n\ne k, \left(n,k\right)\in{\mathcal A}} \theta_{nk} d_{nk}$, $v_n=\theta_{nk}d_{nk}$ for $n\in{\mathcal N}:n\ne k,\left(n,k\right)\in{\mathcal A}$, and $v_n=0$ for $n\in{\mathcal N}:n\ne k,\left(n,k\right)\notin{\mathcal A}$. When deriving equality (c), we use the property that $R_{kt}=R_{tk}$ and $R_{kk}=0$. 

In the result shown in (\ref{appendix:equ:lastlongeqs}), the term $\sum_{\left(i,j\right)\in{\mathcal A}} \theta_{ij} \xi_{ij} \left(\frac{1-c}{2}\right)^2$ is independent of the choice of $k$. We can multiply the rest part of the result by $4$, and get the expression to be maximized in (\ref{equ:theorem:optimalloc}).
\end{proof}

\section{Division of Investigated Area Based on Rides' Origins and Destinations}\label{appendix:sec:mobi:data}
\begin{figure}[t]
  \centering
  \includegraphics[scale=0.47]{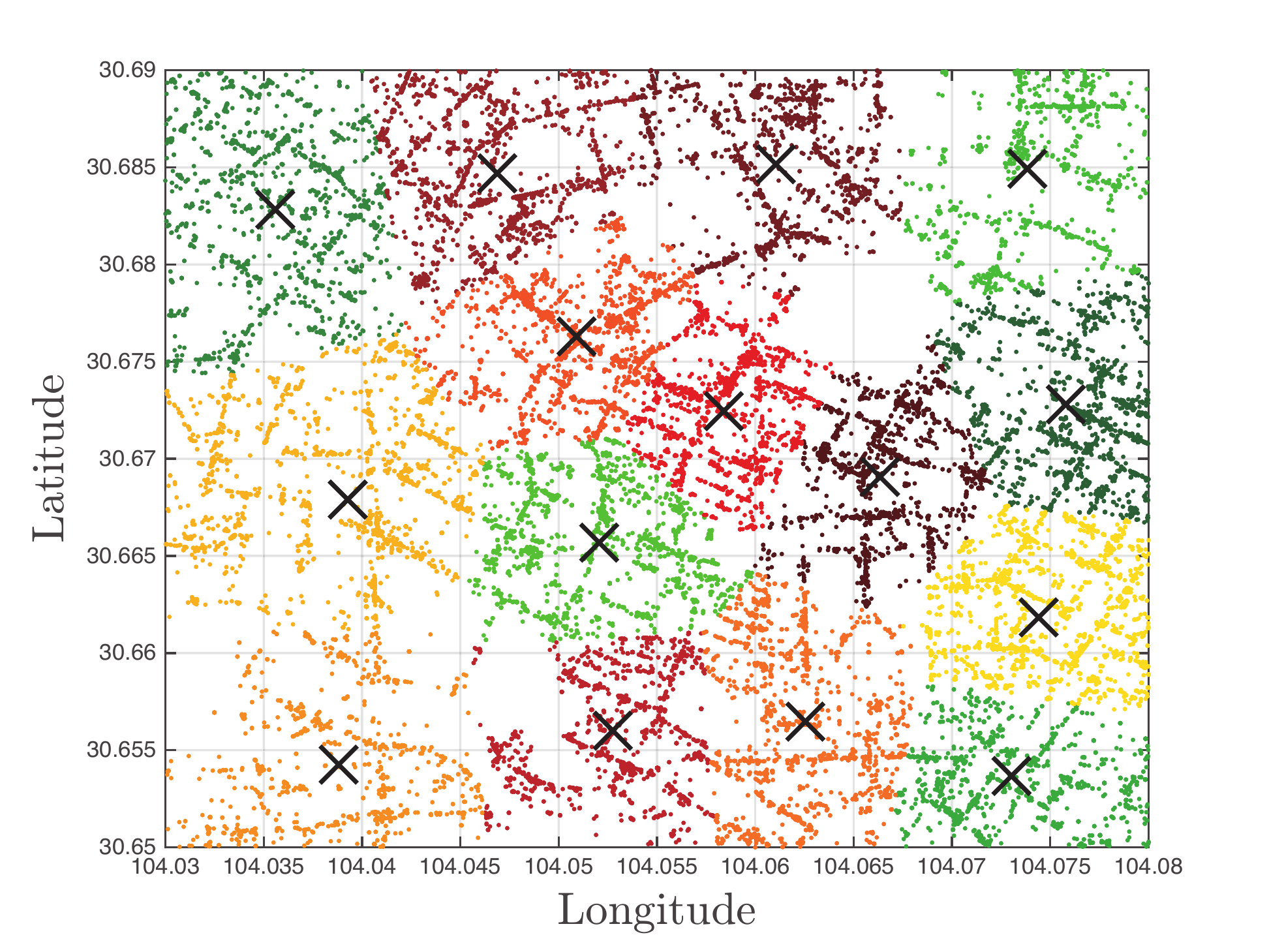}\\
  \caption{Illustration of Clustering.}
  \label{appendix:fig:mobi:2}
\end{figure}

{{In this section, we explain the division of the area based on the distribution of rides' origins and destinations. We use $k$-means clustering to cluster all the origin dots and destination dots into $15$ clusters. We illustrate the clusters in Fig. \ref{appendix:fig:mobi:2}, where different clusters have different colors. The crosses indicate the cluster centroids. We use $15$ locations to represent the $15$ clusters, and let each location's coordinates be the coordinates of the corresponding cluster's centroid.}}

\begin{figure*}
{\small
\begin{align}
\nonumber
& \Pi^{\rm provider} \\
\nonumber
= & \sum_{\left(i,j\right)\in{\mathcal A}} \theta_{ij} \xi_{ij} \left(\frac{1+a_{ij}-c}{2}\right)^2 -\sum_{\left(i,j\right)\in{\mathcal A}} \left(\frac{1}{8}\theta_{ij} \left(1+a_{ij}-c\right)\sum_{n\in{\mathcal N}} \left(R_{jn}-R_{in}\right)v_n\right) \\
\nonumber
 = &\sum_{\left(i,j\right)\in{\mathcal A}:j\ne k} \theta_{ij} \xi_{ij} \left(\frac{1-c}{2}\right)^2 
+ \sum_{\left(i,j\right)\in{\mathcal A}:j=k} \theta_{ij} \xi_{ij} \left(\frac{1+d_{ij}-c}{2}\right)^2\\
\nonumber
& -\sum_{\left(i,j\right)\in{\mathcal A}:j\ne k} \left(\frac{1}{8}\theta_{ij} \left(1-c\right)\sum_{n\in{\mathcal N}} \left(R_{jn}-R_{in}\right)v_n\right)
-\sum_{\left(i,j\right)\in{\mathcal A}:j=k} \left(\frac{1}{8}\theta_{ij} \left(1+d_{ij}-c\right)\sum_{n\in{\mathcal N}} \left(R_{jn}-R_{in}\right)v_n\right)\\
\nonumber
 = &\sum_{\left(i,j\right)\in{\mathcal A}} \theta_{ij} \xi_{ij} \left(\frac{1-c}{2}\right)^2 
+ \sum_{\left(i,j\right)\in{\mathcal A}:j=k} \theta_{ij} \xi_{ij} \left(\left(\frac{d_{ij}}{2}\right)^2+\left(1-c\right)\frac{d_{ij}}{2}\right)\\
\nonumber
& -\sum_{\left(i,j\right)\in{\mathcal A}} \left(\frac{1}{8}\theta_{ij} \left(1-c\right)\sum_{n\in{\mathcal N}} \left(R_{jn}-R_{in}\right)v_n\right)
-\sum_{\left(i,j\right)\in{\mathcal A}:j=k} \left(\frac{1}{8}\theta_{ij} d_{ij}\sum_{n\in{\mathcal N}} \left(R_{jn}-R_{in}\right)v_n\right) \\
\nonumber
\overset{(a)}= &\sum_{\left(i,j\right)\in{\mathcal A}} \theta_{ij} \xi_{ij} \left(\frac{1-c}{2}\right)^2 
\! + \!\!\!\!\! \sum_{\left(i,j\right)\in{\mathcal A}:j=k} \!\!\! \theta_{ij} \xi_{ij} \left(\left(\frac{d_{ij}}{2}\right)^2+\left(1-c\right)\frac{d_{ij}}{2}\right) -\!\!\!\! \sum_{\left(i,j\right)\in{\mathcal A}:j=k} \left(\frac{1}{8}\theta_{ij} d_{ij}\sum_{n\in{\mathcal N}} \left(R_{jn}-R_{in}\right)v_n\right)\\
\nonumber
= &\sum_{\left(i,j\right)\in{\mathcal A}} \theta_{ij} \xi_{ij} \left(\frac{1-c}{2}\right)^2 
+ \sum_{\left(i,j\right)\in{\mathcal A}:j=k} \theta_{ij} \xi_{ij} \left(\left(\frac{d_{ij}}{2}\right)^2+\left(1-c\right)\frac{d_{ij}}{2}\right) \\
\nonumber
& -\!\!\!\!\! \sum_{\left(i,j\right)\in{\mathcal A}:j=k} \left(\frac{1}{8}\theta_{ij} d_{ij}\left(\sum_{n\in{\mathcal N}:n\ne k, \left(n,k\right)\notin{\mathcal A}} \!\!\!\!\!\left(R_{jn}-R_{in}\right)v_n  +\!\!\!\!\! \sum_{n\in{\mathcal N}:n\ne k, \left(n,k\right)\in{\mathcal A}} \!\!\left(R_{jn}-R_{in}\right)v_n +\left(R_{jk}-R_{ik}\right)v_k \right)\right)\\
\nonumber
\overset{(b)}= &\sum_{\left(i,j\right)\in{\mathcal A}} \theta_{ij} \xi_{ij} \left(\frac{1-c}{2}\right)^2 
+ \sum_{\left(i,j\right)\in{\mathcal A}:j=k} \theta_{ij} \xi_{ij} \left(\left(\frac{d_{ij}}{2}\right)^2+\left(1-c\right)\frac{d_{ij}}{2}\right) \\
\nonumber
& -\!\!\!\!\!\! \sum_{\left(i,j\right)\in{\mathcal A}:j=k} \left(\frac{1}{8}\theta_{ij} d_{ij}\left( \sum_{n\in{\mathcal N}:n\ne k, \left(n,k\right)\in{\mathcal A}} \!\!\!\!\!\!\left(R_{jn}-R_{in}\right) \theta_{nk} d_{nk} +\left(R_{jk}-R_{ik}\right)\left(\sum_{n\in{\mathcal N}:n\ne k, \left(n,k\right)\in{\mathcal A}} -\theta_{nk} d_{nk}\right) \right)\right) \\
\nonumber
= &\sum_{\left(i,j\right)\in{\mathcal A}} \theta_{ij} \xi_{ij} \left(\frac{1-c}{2}\right)^2 
+ \frac{1}{4} \sum_{s:\left(s,k\right)\in{\mathcal A}} \theta_{sk} \xi_{sk} \left(d_{sk}^2+2\left(1-c\right)d_{sk}\right) \\
\nonumber
& -\sum_{s:\left(s,k\right)\in{\mathcal A}} \left(\frac{1}{8}\theta_{sk} d_{sk}\left( \sum_{n\in{\mathcal N}:n\ne k, \left(n,k\right)\in{\mathcal A}} \left(R_{kn}-R_{sn}-R_{kk}+R_{sk}\right) \theta_{nk} d_{nk} \right)\right) \\
\nonumber
\overset{(c)}= &\sum_{\left(i,j\right)\in{\mathcal A}} \theta_{ij} \xi_{ij} \left(\frac{1-c}{2}\right)^2 
+ \frac{1}{4} \sum_{s:\left(s,k\right)\in{\mathcal A}} \theta_{sk} \xi_{sk} \left(d_{sk}^2+2\left(1-c\right)d_{sk}\right) \\
\nonumber
& -\sum_{s:\left(s,k\right)\in{\mathcal A}} \left(\frac{1}{8}\theta_{sk} d_{sk}\left( \sum_{t:\left(t,k\right)\in{\mathcal A}} \left(R_{tk}-R_{st}+R_{sk}\right) \theta_{tk} d_{tk} \right)\right) \\
\nonumber
= &\sum_{\left(i,j\right)\in{\mathcal A}} \theta_{ij} \xi_{ij} \left(\frac{1-c}{2}\right)^2 
+ \frac{1}{4} \sum_{s:\left(s,k\right)\in{\mathcal A}} \theta_{sk} \xi_{sk} \left(d_{sk}^2+2\left(1-c\right)d_{sk}\right) \\
& + \frac{1}{4}\sum_{s:\left(s,k\right)\in{\mathcal A}}\sum_{t:\left(t,k\right)\in{\mathcal A}} \frac{1}{2}\theta_{sk} d_{sk}\theta_{tk} d_{tk}  \left(R_{st}-R_{tk}-R_{sk}\right).\label{appendix:equ:lastlongeqs}
\end{align}}
\end{figure*}

\end{document}